\documentclass[preprint]{aastex}
\pdfoutput=1
\usepackage{graphicx}
\usepackage{natbib}
\usepackage{comment}
\begin{document}

\title{Supershells as Molecular Cloud Factories: Parsec Resolution Observations of H{\sc i} and $^{12}$CO(J=1--0) in GSH 287+04--17 and GSH 277+00+36}
\author{J. R. Dawson\altaffilmark{1,2}}
\email{joanne@a.phys.nagoya-u.ac.jp}
\author{N. M. McClure-Griffiths\altaffilmark{2}}
\author{A. Kawamura\altaffilmark{1}}
\author{N. Mizuno\altaffilmark{3}}
\author{T. Onishi\altaffilmark{4}}
\author{A. Mizuno\altaffilmark{5}}
\author{Y. Fukui\altaffilmark{1}}
\altaffiltext{1}{Department of Physics and Astrophysics, Nagoya University, Chikusa-ku, Nagoya, Japan}
\altaffiltext{2}{Australia Telescope National Facility, CSIRO Astronomy \& Space Science, Marsfield NSW 2122, Australia}
\altaffiltext{3}{National Astronomical Observatory of Japan, Mitaka, Tokyo, 181-8588, Japan}
\altaffiltext{4}{Department of Astrophysics, Graduate School of Science, Osaka Prefecture University, 1-1 Gakuen-cho, Nakaku, Sakai, Osaka 599-8531, Japan}
\altaffiltext{5}{Solar-terrestrial Environment Laboratory, Nagoya University, Furo-cho, Chikusa-ku, Nagoya, Aichi 464-8601, Japan}

\begin{abstract}
We present parsec-scale resolution observations of the atomic and molecular ISM in two Galactic supershells, GSH 287+04--17 and GSH 277+00+36. H{\sc i} synthesis images from the Australia Telescope Compact Array are combined with $^{12}$CO(J=1--0) data from the NANTEN telescope to reveal substantial quantities of molecular gas closely associated with both shells. 
These data allow us to confirm an enhanced level of molecularization over the volumes of both objects, providing the first direct observational evidence of increased molecular cloud production due to the influence of supershells. We find that the atomic shell walls are dominated by cold gas with estimated temperatures and densities of $T\sim100$ K and $n_0\sim10$ cm$^{-3}$. Locally, the shells show rich substructure in both tracers, with molecular gas seen elongated along the inner edges of the atomic walls, embedded within H{\sc i} filaments and clouds, or taking the form of small CO clouds at the tips of tapering atomic `fingers'. We discuss these structures in the context of different formation scenarios, suggesting that molecular gas embedded within shell walls is well explained by in-situ formation from the swept up medium, whereas CO seen at the ends of fingers of H{\sc i} may trace remnants of molecular clouds that pre-date the shells. A preliminary assessment of star formation activity within the shells confirms ongoing star formation in the molecular gas of both GSH 287+04--17 and GSH 277+00+36.

\end{abstract}


\section{Introduction}

The neutral interstellar medium (ISM) of the Galactic disk is riddled with 
loops, 
shells and cavities \citep[e.g.][]{heiles79,mcclure02,ehlerova05}. The largest of these structures -- H{\sc i} supershells -- may reach hundreds of parsecs in diameter, with formation energies as high as 
$\sim10^{53}$ ergs. 
Such objects strongly influence the structure and evolution of the Disk ISM. 

The dominant paradigm holds that supershells are formed through the cumulative action of multiple stellar winds and supernovae, which 
blow hot, overpressurized bubbles,
and sweep up the surrounding medium into cool, dense 
shells \citep{bruhweiler80,tomisaka81,mccray87}. The accumulation of the ISM in such superstructures
is one means of 
generating the high densities and column densities 
required for the production of molecular gas, and supershells have long been suggested as drivers of molecular cloud formation \citep[e.g.][]{mccray87,mashchenko94,fukui99,hartmann01}. However, despite a 
substantial
body of theoretical work \citep[e.g.][]{koyama00,bergin04,heitsch08,inoue09}, 
and an growing list of articles documenting supershell-associated molecular gas \citep[e.g.][]{handa86, jung96, fukui99, matsunaga01, yamaguchi01b, dawson08b},
conclusive observational evidence of this phase change occurring in the walls of shells has not yet been found. The degree to which large-scale stellar feedback is driving the production of molecular clouds remains unconstrained. 


Conversely, in a pre-structured and highly inhomogeneous ISM, an expanding supershell will undergo encounters with pre-existing dense clouds, which may compress, disrupt, fragment, or even completely destroy them \citep[e.g.][]{klein94,foster96,mellema02}. The relationship between supershells and the molecular ISM should therefore be governed by the interplay between this potentially destructive process and the creation of new molecular material. 

This paper presents parsec-scale resolution observations of atomic hydrogen and $^{12}$CO(J=1--0) in two Galactic shells, GSH 287+04--17 and GSH 277+00+36, with the aim of investigating the role played by supershells in the evolution of the molecular ISM. These are some of the highest resolution images of any supershell, and the first time a dedicated comparison between the atomic and molecular material in shell walls has been performed.

In the remainder of this introduction we briefly summarize the basic properties of the two shells, before moving on to describe the observational and data reduction techniques in \S\ref{observations}. Sections \ref{results:obschar} and \ref{fitting} discuss the distribution and morphology of the two tracers, and examine the properties of the two phases quantitatively via Gaussian fitting to the shell spectra. Section \ref{coenhancement} investigates the degree of molecularization in the shell volumes, finding evidence of enhanced molecular fractions in both objects, and \S\ref{totalmass} gives their total H{\sc i} and H$_2$ masses. In \S\ref{insitu} and \S\ref{preexisting} we discuss H{\sc i}-CO structures in the shell walls in the context of different formation scenarios. Section \ref{stars} takes a preliminary look at star formation activity in the shell molecular clouds, and our conclusions are finally summarized in \S\ref{conclusions}.

\subsection{GSH 287+04--17: Basic Properties}
\label{gsh287:basics}

GSH 287+04--17, also known as the `Carina Flare' supershell, is a medium-sized, gently expanding Galactic chimney, located in the Sagittarius-Carina Arm at a distance of $2.6\pm0.4$ kpc and Galactocentric radius of $\sim8$ kpc. It was originally discovered in $^{12}$CO(J=1--0) by \citet{fukui99}, who reported a scattering of molecular clouds extending in a wide swath above the Galactic Plane. These clouds showed telltale signs of global expansion, leading the authors to correctly identify them as parts of an expanding superstructure. \citet{dawson08a} later confirmed the existence of the counterpart atomic shell using low resolution $\sim16'$ H{\sc i} data from the Southern Galactic Plane Survey \citep[SGPS;][]{mcclure05}, revealing its global properties and large-scale morphology for the first time. The main body of the shell measures $\sim230\times360$ pc, and it has broken out of the disk at a height of $z\sim280$ pc, with a associated high-latitude emission seen up to heights of $\sim450$ pc above the midplane. The molecular clouds form co-moving parts of the shell, which is estimated to contain total H{\sc i} and H$_2$ masses of $M_{\mathrm{HI}}\sim7\pm3\times10^5~M_{\odot}$ and $M_{\mathrm{H}_2}\sim2.0\pm0.6\times10^5~M_{\odot}$, respectively. The expansion velocity of the shell is $\sim10$ km s$^{-1}$, and its age and formation energy are estimated to be $\sim1\times10^7$ yr and between $0.5-1\times10^{52}$ ergs, based on comparisons with analytical and numerical models. 

\subsection{GSH 277+00+36: Basic Properties}
\label{gsh277:basics}

GSH 277+00+36 is a large outer Galaxy chimney, located at the edge of the Sagittarius-Carina spiral arm at a kinematic distance of $6.5\pm0.9$ kpc and Galactocentric radius of $\sim10$ kpc. It was originally discovered in the low resolution portion of the SGPS \citep{mcclure00}, where it is seen as a prominent void in the bright H{\sc i} emission of the Plane. With a main body $\sim610$ pc in diameter and chimney extensions that reach more than 1 kpc above the midplane, this large and evolved supershell has swept up an estimated atomic mass of $M_{\mathrm{HI}}\sim3\pm1\times10^6~M_{\odot}$, and is expanding with a velocity of $\sim20$ km s$^{-1}$. The age and formation energy of the shell are estimated to be between $1-2\times10^7$ yr and $0.9-2.4\times10^{53}$ ergs, where the higher values in these ranges are taken directly from \citet{mcclure00} and the lower values are derived using the same analytical formula as \citet{dawson08a} in order to better illustrate the difference between the two shells. High-resolution ($\sim3'$) H{\sc i} observations by \citet{mcclure03} revealed narrow, well-defined walls, sharply delineated along their inner edges, with a great deal of complex substructure including knots, filaments and `drips'. No molecular observations have been previously reported for this shell.

\section{Observations}
\label{observations}

\subsection{HI Observations and Reduction}
\label{hiobs}

This paper makes use of new Australia Telescope Compact Array (ATCA) mosaicing observations of GSH 287+04--17 and previous observations of GSH 277+00+36 from \citet{mcclure03}. Here, the GSH 277+00+36 mosaicing observations are reprocessed completely from the original raw data files to ensure consistency in the data reduction procedures for both shells.

GSH 287+04--17 was observed over two sessions in June and November of 2007 with the EW352 and EW367 array configurations. These configurations provided nearly uniform baseline coverage from 31 to 367 m at intervals of approximately 15 m. The observations covered an 81 deg$^2$ region between $283 < l < 292^{\circ}$ and $1.5 < b < 10.5^{\circ}$, in 653 pointings arranged in a hexagonal pattern at a spacing of $23\arcmin$. Although this value is slightly less than Nyquist spacing for the ATCA $34\arcmin$ primary beam, the sensitivity function only varies by $\sim2.5\%$ between pointings, and the smaller number of pointings allows more snapshots per pointing to improve $u$-$v$ coverage. A total of fifteen 40 s integrations were made for each pointing at widely spaced hour angles, and the standard sources PKS 1934-638 and PKS 1039-47 were observed for bandpass and absolute flux-density calibration, and amplitude and phase calibration respectively. The correlator configuration provided a velocity channel width of 0.41 km s$^{-1}$ at 1.420 GHz. 

GSH 277+00+36 was observed in three sessions between October 2001 and March 2002. The observing strategy for GSH 277+00+36 differed from that of GSH 287+04--17 in the following points: 1. A total of 841 pointings were made covering the regions $272 < l < 284^{\circ}$, $1.0 < b < 5.5^{\circ}$ and $272 < l < 284^{\circ}$, $-7.0 < b < -1.0^{\circ}$; 2. each pointing was observed in eight 60 s snapshots; 3. the standard sources used for amplitude and phase calibration were PKS B1039-47 and PKS B0843-54; 4. the correlator configuration provided a velocity resolution of 0.82 km s$^{-1}$ at 1.420 GHz.

For both supershells, coverage of the Galactic Plane region ($|b| < 1.0^{\circ}$) was provided by ATCA data taken during the Southern Galactic Plane Survey \citep[SGPS,][]{mcclure05}. Pointings were spaced $19\arcmin$ apart in a hexagonal configuration and a total of forty 30 s integrations were made towards each point. The array configurations used were 210, 375, 750A, 750B, 750C, and 750D, with PKS B1934-638 observed for bandpass and absolute flux calibration and PKS 0823-500 observed for phase and amplitude calibration. Further details may be found in \citet{mcclure05}. 

Calibration and imaging were performed using standard routines from the MIRIAD software package \citep{miriadguide}. Calibration, baseline subtraction and Doppler correction were carried out in the $u$-$v$ domain. The individual pointings were then linearly combined and imaged using the MIRIAD task INVERT -- a standard grid and fast Fourier transform technique -- with superuniform weighting. Deconvolution was performed using the maximum entropy-based deconvolution algorithm MOSMEM \citep{sault96}, and the images were restored using the task RESTOR. For both shells the dimensions and position angle of the dirty beam varied somewhat with the pointing. For this reason we chose to restore with a circular gaussian beam whose FWHM matched the largest values measured for the major axes of the dirty beams. This corresponds to $2.5\arcmin$ for GSH 287+04--17 and $3.0\arcmin$ for GSH 277+00+36.

The synthesis images for both supershells were then combined with Parkes telescope single dish data from the Galactic All-Sky Survey \citep[GASS, ][]{mcclure09,kalberla10}. GASS has an effective angular resolution of $\sim16\arcmin$, an effective velocity resolution of 1.0 km s$^{-1}$, and a brightness temperature sensitivity of $\sim60$ mK. The GASS data were re-gridded to match the ATCA data, and the two were then linearly combined in the Fourier domain using the MIRIAD task IMMERGE. 

The final synthesized beam sizes are $2.5\arcmin$ and $3.0\arcmin$ for GSH 287+04--17 and GSH 277+00+36 respectively, and the velocity resolution of both datasets is 1.0 km s$^{-1}$ (although the channel width of the gridded data is 0.82 km s$^{-1}$). The r.m.s. noise in a single 0.82 km s$^{-1}$ channel is $\sim1$ K for both objects. 

\subsection{$^{12}$CO(J=1--0) Observations and Reduction}
\label{coobs}

$^{12}$CO(J=1--0) observations of the GSH 287+04--17 region were made in two sessions from April to November 1998 and November 2001 to May 2002, using the NANTEN 4m telescope. Partial or complete datasets have previously been published in several studies \citep{fukui99, matsunagaphd, dawson08a}, and most recently have been presented in catalogue form by \citet{dawson08b}. The telescope half power beam width at 115 GHz is $\sim2.6\arcmin$. All observations were made by position switching, with pointing centers arranged in a square grid at spacings of $2\arcmin$ or $4\arcmin$. The vast majority of supershell-associated CO emission ($\sim99\%$ of the total integrated intensity) is covered at $2\arcmin$ spacing. Further details of the observing strategy and region coverage may be found in \citet{dawson08b}.

Intensity calibration was made by the chopper wheel method, and Orion KL ($\alpha_{B1950}=5^h32^m47.5^s$, $\delta_{B1950}=-5^{\circ}24\arcmin21\arcsec$) was observed as a standard calibrator source. The 2048 channel acousto-optical spectrograph provided a velocity coverage and resolution of 100 km s$^{-1}$ and 0.1 km s$^{-1}$, respectively. Typical system noise temperatures at 115 GHz were $\sim$250 K, resulting in an r.m.s. noise of $\sim0.5$ K per channel.

The GSH 277+00+36 region was observed during the late 1990s as part of a pilot run for the NANTEN Galactic Plane Survey \citep{mizuno04}. The data cover a Galactic latitude range of $|b|<5^{\circ}$ between $272 < l < 280^{\circ}$ and $-4 < b < +3^{\circ}$ between $280 < l < 284^{\circ}$, with velocity coverages of $-100 < v_{lsr} < 100$ km s$^{-1}$ and $-50 < v_{lsr} < 50$ km s$^{-1}$ for these two regions respectively. The observations were made by position switching at a grid spacing of $4\arcmin$, at a velocity resolution of 0.65 km s$^{-1}$. The final data cube is gridded to a resolution of 1.0 km s$^{-1}$. The r.m.s. noise per channel is typically $\sim0.35$ K. 

The differing data quality in the two supershell regions necessitates further brief comment. In GSH 287+04--17 the detection limit in $L_{\mathrm{CO}}$ is estimated to be $\sim3.5$ pc$^2$ K km s$^{-1}$ \citep{dawson08b}, where $L_{\mathrm{CO}}$ is the sum of the velocity-integrated intensities over the entire projected area of a cloud. For GSH 277+00+36, the lower sensitivity, wider grid spacing and larger distance, combined with non-Nyquist noise arising from poor baseline stability, conspire to produce a detection limit approximately an order of magnitude higher, which affects our ability to reliably detect small or low-brightness clouds.

Section \ref{coenhancement} briefly makes use of NANTEN Galactic Plane Survey datacubes covering regions immediately adjacent to the two shells. This additional data covers the regions $284 < l < 293^{\circ}$, $-4 < b < +2^{\circ}$ for GSH 287+04--17, and $270 < l < 272^{\circ}$, $-5 < b < +4^{\circ}$ for GSH 277+00+36. The former is a composite of observations made at grid spacings of either $2\arcmin$ or $4\arcmin$, at a velocity resolution of 0.1 km s$^{-1}$, with r.m.s. noise levels per channel ranging between $\sim0.3$ and $\sim0.9$ K. The latter was made at a grid spacing of $8\arcmin$, and has a velocity resolution of 1.0 km s$^{-1}$, with an r.m.s. noise of $\sim0.2$ K per channel.

\section{Results} 
\label{results}

\subsection{Observational Characteristics of H{\sc i} and CO in the Two Shells}
\label{results:obschar}

\subsubsection{GSH 287+04--17}
\label{results:287}

Figure \ref{fig:287chs} shows velocity channel maps of the H{\sc i} and CO(J=1--0) emission in the GSH 287+04--17 region, both at a spatial resolution of $\sim 2$ pc. Each image displays an average over three H{\sc i} velocity channels, corresponding to a width of 2.46 km s$^{-1}$. These new high-resolution H{\sc i} data reveal a wealth of clumpy, knotted and filamentary substructure in the atomic shell walls, on size-scales ranging down to the resolution limit. The global morphology of the supershell has been described in detail by \citet{dawson08a} and we do not duplicate their discussion here. Instead we focus on the information this matched-resolution H{\sc i} and CO data provides on the relationship between the atomic and molecular medium on local scales. 

At velocities below $\sim-20$ km s$^{-1}$ (panels \textit{a} to \textit{f} in figure \ref{fig:287chs}) velocity crowding is minimal, and emission belonging to the supershell is clear and unambiguous. The atomic and molecular ISM form striking configurations. These include cases in which molecular gas is elongated along the inside of the atomic shell wall, neatly embedded within H{\sc i} filaments or extended clouds, and numerous cases of small CO clouds at the tips of tapering H{\sc i} features, whose morphology suggests sculpting of the atomic medium by the shell energy source.

Figure \ref{fig:287regions} shows a closeup of some of the most prominent CO-H{\sc i} features in the datacube. Panel \textit{a1} shows a section of the approaching wall; hereon referred to as the `approaching limb complex'. Here, tapered fingers of H{\sc i} gas point radially inwards towards the shell interior, some tipped with clumps of CO-bright molecular gas. The shell walls between these features form arch-like structures, with a typical separation of $\sim20-40$ pc. The CO clouds are small, with estimated molecular masses of $70\lesssim M_{\mathrm{H}_2}\lesssim 600~M_{\odot}$ \citep{dawson08b}. In contrast, the CO-H{\sc i} complex at $(l, b)\approx(286^\circ, 5^\circ)$ shows a bright, elongated molecular cloud apparently embedded within the curved H{\sc i} wall. Here, the CO and H{\sc i} distributions are tightly correlated, to the extent that thin, twisted filaments within the wall are in places neatly traced in both transitions (shown in panel \textit{a2} of the figure). The mass of the largest CO cloud within this complex is estimated to be $M_{\mathrm{H}_2}\sim5500~M_{\odot}$, and it contains an active star forming region at $(l, b)\approx(285.9^\circ, 4.5^\circ)$ \citep[][see also \S\ref{stars}]{dawson07}.

Panel \textit{b} of the same figure shows emission originating from the high latitude regions of the supershell. The prominent group of  CO clouds in this region are located $z\sim400$ pc above the Galactic midplane. This is $\sim8$ times higher than the molecular gas scale height at the Galactocentric radius of the shell \citep[$\sigma_z\sim50$ pc at $R\sim8$ kpc;][]{clemens88,malhotra94a} and illustrates the likely importance of supershells in providing dense material to the Disk-Halo interface. This region, hereafter referred to as the `high latitude complex', is dominated by a large, bright V-shaped atomic structure extending over 200 pc in length, whose foremost regions contain an estimated $\sim1.1\times10^4~M_{\odot}$ of molecular hydrogen. The leading CO cloud in this group is located at the very tip the atomic feature, offset from the bulk of the H{\sc i} emission. In contrast, the remainder of the molecular gas in the region is generally well associated with bright knots and filaments in the atomic medium. To the left of this complex, a single isolated molecular cloud at $(l, b)\approx(290.3^\circ, 7.6^\circ)$ is seen at the very end of a long, filamentary arc of H{\sc i}.

At velocities greater than $-20$ km s$^{-1}$ velocity crowding begins to become problematic, and it is not always possible to disentangle shell emission from the bright and complex H{\sc i} background. Nevertheless, some associated structures are evident, and show similar characteristics to the features described above. Panel \textit{c} shows a complex identified as part of the shell's receding limb, which contains several of the brightest molecular clouds in the sample. These clouds occupy positions along the lower edges of the H{\sc i} feature, on the side facing the Galactic plane, with a single cloud at $(l, b)\approx(287.1^\circ, 2.4^\circ)$ offset towards the tip of a protrusion of atomic gas. The largest have estimated H$_2$ masses of between several hundred and $\sim2000~M_{\odot}$, and show signs of active star formation (see \S\ref{stars}).

\subsubsection{GSH 277+00+36}
\label{results:277}

Our CO(J=1--0) data newly reveal a population of molecular clouds scattered throughout the walls of GSH 277+00+36. The global properties and detailed morphology of the atomic shell have been described previously by \citet{mcclure00, mcclure03}, the latter using the same high-resolution H{\sc i} observations as the present work. However, the comparison with molecular tracers is entirely new, and sheds the first light on the hitherto overlooked molecular component of this evolved outer Galaxy shell. All H$_2$ masses quoted in this section are derived using the X-factor as described in \S\ref{fitting:coldens}. 

Figure \ref{fig:277chs} shows velocity channel maps of the H{\sc i} and CO(J=1--0) emission in GSH 277+00+36, at spatial resolutions of $\sim6$ pc and $\sim7$ pc, respectively. Each image displays an average over three H{\sc i} velocity channels, corresponding to a width of 2.46 km s$^{-1}$. Molecular clouds are scattered throughout the main walls, forming co-moving parts of the H{\sc i} shell, with the clearest associations seen around the systemic velocity (panels \textit{e} to \textit{j}). Despite differences in age, distance and evolutionary stage, features in the shell walls show notable similarities to GSH 287+04--17, with CO clouds found embedded within the main walls, offset towards the tips of H{\sc i} features, and elevated to high altitudes above the midplane.

Figure \ref{fig:277regions}, panel \textit{a1} shows a closeup of a section of `scalloped' H{\sc i} wall, in which the atomic walls form arcs and `fingers' similar to those seen in the approaching limb of GSH 287+04--17 (panel \textit{a1} of figure \ref{fig:287regions}). A molecular cloud of $M_{\mathrm{H}_2}\sim1000~M_{\odot}$ is offset towards the tip of one of these atomic protrusions. To the right of this feature CO emission is detected in a single position emanating from the very tip of a smaller, narrower `drip' of H{\sc i} protruding from the shell wall (panel \textit{a2}). This small, unresolved CO cloud is on the very limit of detectability for the dataset, permitting the possibility that other similar atomic `drips' on the inner walls of the shell could harbor as-yet-undetected molecular components. This region is hereon referred to as the `molecular drip' region. 

Panel \textit{b} of the same figure shows a large, bright molecular cloud complex embedded within the main shell walls. This prominent group of CO clouds spans $\sim100$ pc, with an estimated molecular mass of $M_{\mathrm{H}_2}\sim5\times10^4~M_{\odot}$. Its rightmost member contains the giant H{\sc ii} region RCW 42. 

Panel \textit{c} shows a large ($\sim5000~M_{\odot}$) molecular cloud associated with the walls of one of the shell's chimney conduits. Located $z\sim450$ pc below the Galactic midplane, its altitude is $\sim6$ times the molecular gas scale height in the outer Galaxy \citep[$\sigma_z\sim80$ pc;][]{heyer98}, providing another example of a massive high-latitude molecular cloud associated with a known supershell. 

\subsubsection{Morphology Classes: Embedded and Offset CO Clouds}
\label{results:morph}

The discussion above refers to `embedded' CO clouds, that appear to be well embedded in parts of the atomic shell, and `offset' CO clouds, that are found at the tips of H{\sc i} features, offset from the bulk of the atomic material. In the analysis to follow we find it useful to explicitly distinguish between these two loose morphological categories. Those clouds treated as `embedded' and `offset' are marked in figures \ref{fig:287regions} and \ref{fig:277regions}. 

\subsection{Quantitative Properties from Gaussian Decomposition}
\label{fitting}

Gaussian decomposition of H{\sc i} spectra enables the separation shell emission from the multi-component background, and hence the extraction of accurate line profiles from which to obtain quantitative information on the atomic shell ISM. Equally importantly, it enables meaningful comparisons with the $^{12}$CO(J=1--0) line, allowing us to examine how H{\sc i} properties may vary according to the presence and characteristics of molecular gas.

Here Gaussian fitting is performed for a sub-sample of spectra from the shells. We select the most prominent H{\sc i}-CO features in the approaching limb and high latitude complexes in GSH 287+04--17 (panels \textit{a} and \textit{b} in Figure \ref{fig:287regions}), and the GSH 277+00+36 molecular `drip' region (panel \textit{a} of Figure \ref{fig:277regions}). Shell emission from these regions is generally distinct and unconfused, making them well suited to this quantitative analysis.
The sample of observed spectra in the GSH 277+00+36 region is small, and is included mainly to check for consistency between the two shells. 

\subsubsection{Implementation}

The H{\sc i} and CO datacubes are re-gridded to $2'$ and $3'$ spatial grids for the GSH 287+04--17 and GSH 277+00+36 regions respectively. The fitting is implemented interactively in IDL using the non-linear least squares fitting package MPFIT \citep{markwardt09}, with initial guesses for the model function and component parameters either made by eye or based on the fit parameters for neighboring spatial points. Examples of fitted spectra are shown in figure \ref{fig:examplespecs}.

In cases where the signal component from the shell appears to be a blending of more than one Gaussian, the statistical F-test is applied to determine whether the decrease in $\chi^2$ that arises from the addition of the extra component is statistically significant \cite[e.g.][]{westmoquette07}.
We require this decrease to be significant at the 4 sigma level for the double fit to be retained. Where such double components are present they are labelled according to their velocities, with component 1 designated as the more blueshifted of the two lines. Post-processing we have become aware of work that cautions against this application of the F-test, demonstrating that it is technically statistically incorrect despite its extensive use within the community \citep{protassov02}. Nevertheless, tests on synthetic spectra demonstrate that in our case this approach reliably identifies double components, provided that their separation is comparable to or greater than the velocity resolution of the data. 

Because we are interested primarily in accurate background subtraction, and because we have deliberately chosen regions where the signal emission is well defined, the fitting is free from many of the difficulties that can often plague multi-component Gaussian decomposition \citep[see e.g.][]{haud00}. Although in some cases several solutions may fit the profiles similarly well, or the routine may fail to locate the true minimum $\chi^2$ fit for a given model function, in practice 
the exact choice of background fit does not greatly affect the parameters of the signal component. In the vast majority of cases the variation in signal parameters is $< 10\%$, and never greater than $\sim20\%$ even for the most ambiguous spectra in our sample. This level of accuracy is sufficient for our needs.

\subsubsection{Column Densities and Number Densities}
\label{fitting:coldens}

Atomic and molecular hydrogen column densities through the shell walls are estimated from the fit solutions. $N_\mathrm{HI}$ takes values of $1-8\times10^{20}$ cm$^{-2}$, and  $N_{\mathrm{H}_2}$ of $1-60\times10^{20}$ cm$^{-2}$. At all points where molecular gas is observed, $N_{\mathrm{H}_2} \gtrsim N_\mathrm{HI}$.

Here, $N_\mathrm{HI}$ is estimated from the brightness-column density relation, $N_\mathrm{HI}=1.8\times10^{18}\int{T_{b}(\mathrm{HI})\ dv}$ \citep[e.g.][]{dickey90}, which holds in the optically thin regime and returns a lower limit if optical depth is not negligibly small. $N_{\mathrm{H}_2}$ is estimated using the empirical `X-factor' relating $^{12}$CO(J=1--0) integrated intensity and H$_2$ column density, $N_{\mathrm{H}_2}=X\int{T_{b}(\mathrm{CO})\ dv}$, where we adopt a value of $X = 1.6\times10^{20}$ K$^{-1}$ km s$^{-1}$ cm$^{-2}$ \citep{hunter97}. 

Figure \ref{fig:nhimaps} shows H{\sc i} column density, $N_\mathrm{HI}$, across the three fitted regions, as well as the ratio $N_{\mathrm{H}_2}/N_\mathrm{HI}$ in regions where molecular gas is detected. Figure \ref{fig:nhihist} shows histograms of $N_\mathrm{HI}$, color-coded by the presence or absence of spatially coincident CO emission.

Embedded CO clouds are located in zones of high atomic column density. In contrast, offset clouds are associated with very low values of $N_\mathrm{HI}$. This trend is seen clearly in figure \ref{fig:nhihist}, where the two classes of cloud populate distinct regions within the histograms. It is also reflected in the high $N_{\mathrm{H}_2}/N_\mathrm{HI}$ ratios seen towards offset CO clouds, which peak at $\sim 10$, $\sim 30$ and $\sim6$ in the approaching limb complex, high latitude complex, and molecular drip region respectively. This quantitative trend is consistent with the morphological criteria by which the cloud classes were originally defined (see \S\ref{results:morph}). 

By assuming a depth roughly equal to the characteristic size a feature subtends on the sky, a crude estimate of H{\sc i} and H$_2$ number densities in the shell walls may be obtained. This yields $n_\mathrm{HI}\sim10$ cm$^{-3}$ and $n_{\mathrm{H}_2}\sim10$--100 cm$^{-3}$ for all three regions. The former is consistent with the densities expected of atomic CNM. The latter is somewhat low for molecular clouds, although this is likely to be due in part to low fillings factors within the NANTEN beam.

\subsubsection{Velocity Offsets Between Atomic and Molecular Emission}

H{\sc i} and CO intensity-weighted mean velocities, $v_{0,\mathrm{HI}}$ and $v_{0,\mathrm{CO}}$, provide information on the bulk motions of the atomic and molecular gas. Figure \ref{fig:v0maps} shows images of $v_{0,\mathrm{HI}}$, as well as the velocity difference $(v_{0,\mathrm{CO}} - v_{0,\mathrm{HI}})$. 
The uncertainty on this velocity difference is small across both GSH 287+04--17 regions, where it is estimated to be strictly $<0.2$ km s$^{-1}$. However, for GSH 277+00+36 the lower CO data quality leads to much larger uncertainties of $\sim1$ km s$^{-1}$, and this region is therefore ignored in the analysis below.

No systematic offset is found between the intensity-weighted mean velocities of the two tracers. The mean values of $(v_{0,\mathrm{CO}} - v_{0,\mathrm{HI}})$ averaged over all CO-detected points are consistent with zero for both the approaching limb and high latitude complexes. Nevertheless, there is considerable local variation on scales of a few parsecs, with $(v_{0,\mathrm{CO}} - v_{0,\mathrm{HI}})$ taking values of up to $\pm2.5$ km s$^{-1}$. The largest of these local velocity differences are observed in embedded CO clouds, and appear to arise where CO is either associated with one of an H{\sc i} double component pair or offset towards one side of a wide atomic line. In contrast, $v_{0,\mathrm{CO}}$ and $v_{0,\mathrm{HI}}$ in offset clouds mostly agree to within $\sim\pm0.5$ km s$^{-1}$. For many points this separation is on the borderline of statistical significance.

With the exception of locally-driven expansion around the H{\sc ii} region at $(l,b)=(285.9, 4.5)$, we are unable to identify any systematic patterns in the H{\sc i}-CO velocity offset that might indicate coherent macroscopic gas motions such as infall onto individual clouds.

\subsubsection{Velocity Dispersion and Kinetic Temperature}
\label{fitting:deltav}

The velocity dispersion of the H{\sc i} gas provides constraints on its kinetic temperature. Under the assumption of pure thermal line broadening, $T_k$ is 
determined directly from the H{\sc i} Gaussian FWHM, $\Delta v$, meaning that linewidth can be used to obtain a strict upper limit on kinetic temperature, $T_{k,max}=21.9~\Delta v^2$. 
In the following, $\Delta v$ is defined as the intensity-weighted velocity dispersion, $\sigma_v$, multiplied by the factor $\sqrt{8ln(2)}$. This corresponds directly to the Gaussian FWHM in the case of a single component fit, and gives an equivalent combined weighted velocity dispersion when applied to a multiple-component fit. 

Figure \ref{fig:lwmaps} shows images of the total FWHM velocity dispersion of the atomic shell emission, $\Delta v_{\mathrm{HI}, tot}$, as well as the values for individual Gaussian components, $\Delta v_{\mathrm{HI}, 1}$ and $\Delta v_{\mathrm{HI}, 2}$, where present. Figure \ref{fig:lwhist} shows histograms of the same quantities, color-coded by the presence or absence of spatially coincident CO emission. 

Velocity dispersions are very small; as low as $\Delta v_{\mathrm{HI}, tot}\sim2$ km s$^{-1}$ in the narrowest profiles. Mean values are 4.4, 4.0 and 4.8 km s$^{-1}$ in the approaching limb complex, the high latitude complex, and the molecular drip region, respectively. Many of the broadest profiles arise from the blending of two narrower components. This is illustrated by the fact that replacing the intensity-weighted velocity width in blended profiles with $\Delta v_{\mathrm{HI}, 1}$ or $\Delta v_{\mathrm{HI}, 2}$ results in a reduction in these means by $\sim0.4$ km s$^{-1}$. We also expect there to be more positions where blended profiles remain unresolved in our analysis.

These narrow linewidths indicate cold H{\sc i}. For pure thermal broadening, 4 km s$^{-1}$ corresponds to a kinetic temperature of only 350 K. 
In practice, turbulence also contributes significantly to the observed linewidth, and the true value of $T_k$ is considerably lower. \citet{heiles03} find that for the CNM in the solar neighborhood, turbulence and thermal motions generally make comparable contributions to $\Delta v$. Their data is best fit by the following expression: $\log~T_{k,max}=1.32\log~T_s - 0.11$, where 
$T_s$, the spin temperature, is equal to the true value of $T_k$ for the case of the CNM.
Applying this relation to our data yields a mean $T_k$ of roughly
$\sim100$ K for all three sample sections of shell wall. In non-fitted regions too, there is evidence of similarly narrow linewidths throughout both GSH 287+04--17 and GSH 277+00+36. Taken together with the high H{\sc i} number densities derived above, these linewidths provide robust evidence of shell walls that are dominated by cold gas. 

Comparing molecular and non-molecular regions, we find no statistically significant global
correlation between H{\sc i} velocity dispersion and the presence of CO. Nevertheless, significant differences are observed in $\Delta v_{\mathrm{HI},tot}$ towards the two classes of CO cloud. Offset clouds are associated exclusively with gas with $\Delta v_{\mathrm{HI},tot} < 4$ km s$^{-1}$, whereas embedded clouds tend to be found towards gas with $\Delta v_{\mathrm{HI},tot} > 4$ km s$^{-1}$. This trend reflects a correlation between the presence of embedded CO and multiple blended components in H{\sc i} line profiles. 

Finally, 
CO linewidths are on average narrower than H{\sc i}, with mean values of $\Delta v_{\mathrm{CO},tot}$ of 2.0 and 1.9 km s$^{-1}$. 
A weak correlation between $\Delta v_{\mathrm{HI},tot}$ and $\Delta v_{\mathrm{CO},tot}$ is found, with Pearson's correlation coefficients of $\sim0.5$ and $\sim0.3$ for the approaching limb and high latitude complexes respectively. This trend reflects a preference for wide or blended CO profiles in regions where H{\sc i} profiles are also wide or blended, and suggests that CO is not strictly localized to single narrow zones within large columns of H{\sc i}, but exhibits a more complex spatial distribution.

\subsection{Enhanced Molecular Fraction in the Shells} 
\label{coenhancement}

A critical question for supershell-driven molecular cloud formation is whether we observe
an enhanced degree of molecularization in the volumes affected by 
the shells. We here adopt the ratio of total CO to total H{\sc i} integrated intensity, $\sum I_{\mathrm{CO}}/\sum I_{\mathrm{HI}}$, 
as a measure of the molecular fraction, since both quantities scale approximately linearly with mass. 
The value of this ratio within the volumes affected by our shells is then compared with that in quiescent `background' regions in which no recent supershell activity has taken place. 

Low resolution ($\sim15\arcmin$) GASS H{\sc i} datacubes are used, since it is necessary to expand the data coverage to include regions outside those observed with the ATCA. Because we are concerned with integrated intensities summed over large $l$-$b$-$v$ volumes the loss of resolution is of no concern. For CO, we make use of available NANTEN GPS data (see \S\ref{coobs}). 

Both shells occupy coherent volumes in $l$-$b$-$v$ space, and are seen in all three planes as a characteristic void in H{\sc i}, often surrounded by a brightened annulus or clumps. We work in the $l$-$v$ plane, defining the extent of the shells by eye at regular latitude intervals throughout the cube. These 2D shell masks are then assembled into a 3D volume enclosing all emission from both the shell and the void. All remaining emission in the cube is designated as background. 

There are a number of environmental factors that significantly affect molecular gas abundances; most notably altitude above the Galactic midplane, Galactocentric radius, and proximity to spiral arms. 
The spatio-velocity dimensions of each masked cube must therefore be carefully trimmed
to ensure that the non-shell background regions match as closely as possible the Galactic environment in which the shell is found. In practice this involves the following: 1. excluding all emission outside the latitude ranges of the shells; 2. excluding velocities containing contributions from local emission; 3. avoiding the tangent of the Carina Arm (which affects the vicinity around both shells to some extent); 4. ensuring that where variation in distance and Galactocentric radius is unavoidable, this variation is not only minimal, but also well behaved across the cube and well represented in both shell and background regions. The ratio $\sum I_{\mathrm{CO}}/\sum I_{\mathrm{HI}}$ is then derived for the shell and background regions by summing the CO and H{\sc i} integrated intensities contained in each. 
 
The results indicate a significant enhancement in the molecular fraction in both shells. Figure \ref{fig:coenhancement} plots the variation in $\sum I_{\mathrm{CO}}/\sum I_{\mathrm{HI}}$ with Galactic latitude 
both in and out of the two shells, illustrating that an overabundance of CO is consistently observed throughout almost the entire extent of both objects. For GSH 287+04--17 we find a total in-shell ratio of $[\sum I_{\mathrm{CO}}/\sum I_{\mathrm{HI}}]_{sh}=1.3\pm0.4\times10^{-3}$ compared to a background ratio of $[\sum I_{\mathrm{CO}}/\sum I_{\mathrm{HI}}]_{bg}=0.5\pm0.3\times10^{-3}$. For GSH 277+00+36 the equivalent figures are $3.3\pm0.8\times10^{-4}$ and $0.9\pm0.3\times10^{-4}$, implying an enhancement in molecular fraction by a factor of $\sim3$ for both shells. 

The uncertainties on the above figures are estimated via experimentation with a wide variety of different shell masks and spatio-velocity ranges (including some masks defined in the $l$-$b$ plane). A statistically significant increase of $\sum I_{\mathrm{CO}}/\sum I_{\mathrm{HI}}$ is consistently measured across all tested configurations, reflecting a genuine concentration of CO emission in volumes affected by the shells. This enhancement spans values of between around 2.0 and 5.5.

A similar enhancement in mean per-pixel CO intensity is also confirmed over the volumes of both shells. While the non-trivial conversion between velocity and distance means that this quantity cannot readily be translated into a volume density, it nevertheless demonstrates a genuine increase in CO brightness within the two shell regions. This is important, because it demonstrates that the enhanced molecular fraction obtained above reflects a genuine over-abundance of CO and does not arise from H{\sc i} integrated intensity significantly underestimating atomic mass in the CNM-rich gas.  

This result provides clear and direct observational support for supershell-driven molecular cloud formation. Crudely, we suggest that as much as 2/3 of the CO emission in both GSH 287+04--17 and GSH 277+00+36 may arise as a direct result of the influence of these shells. If this result is found to hold true on Galactic scales, it will have profound implications for our understanding of the role played by stellar feedback on the evolution of the molecular phase.

\subsection{Total Atomic and Molecular Masses}
\label{totalmass}

It is appropriate to comment briefly on the total swept-up masses of the two shells, since the molecular mass of GSH 277+00+36 is newly determined in this work. From the analysis above, the total molecular mass of GSH 277+00+36 is estimated to be $M_{\mathrm{H}_2}\sim2.1\pm0.6\times10^5~M_{\odot}$, where the error estimate arises from uncertainties on the kinematic distance. The atomic mass is estimated by \citet{mcclure00} as $M_{\mathrm{HI}}\sim3\pm1\times10^6~M_{\odot}$. For comparison, \citet{dawson08a} estimate the total H{\sc i} and H$_2$ masses of GSH 287+04--17 to be $M_{\mathrm{HI}}\sim7\pm3\times10^5~M_{\odot}$ and $M_{\mathrm{H}_2}\sim2.0\pm0.6\times10^5~M_{\odot}$, respectively. These figures imply a lower degree of molecularization in GSH 277+00+36. This is consistent with its location in the outer Galaxy, where ambient densities are generally lower. Indeed, the implied initial mean number densities of the pre-shell medium are $n_0\sim1$ cm$^{-1}$ and $n_0\sim3$ cm$^{-1}$ for GSH 277+00+36 and GSH 287+04--17, respectively. We note that these absolute molecular fractions are distinct from the level of \textit{enhancement} in the molecular fraction in the shell volumes that was computed above.

\section{Discussion}
\label{discussion}

Both GSH 287+04--17 and GSH 277+00+36 are CNM-rich, with significant masses of associated molecular gas. Crucially, the degree of molecularization in the shell volumes appears to be higher than in equivalent quiescent regions, strongly arguing for an enhanced level of molecular gas production due to the influence of the shells. Yet in an inhomogeneous ISM the ambient medium is pre-structured; containing zones of cold, dense material whose origin predates the supershells. Molecular gas may be formed in-situ from the swept-up medium, but this process must occur in tandem with the interaction of the shells with pre-existing molecular clouds. 

The embedded and offset CO clouds identified in \S\ref{results:obschar} provide some of the best candidates for these two processes. Embedded clouds -- with their close association with the material of the main shell walls -- are strongly suggestive of molecular gas that has condensed from the swept-up atomic medium. In contrast, offset clouds -- located at the tips of swept-back H{\sc i} features -- may be well explained as the remnants of an initial medium that was peppered with dense clumps.

The remainder of this paper demonstrates the plausibility of these two scenarios for archetypal examples of embedded and offset clouds. We use these objects as a basis from which to explore some key issues surrounding molecular cloud formation and destruction, focussing particularly on cloud sweep-up/formation timescales, the outcome of shock-cloud interactions, and the requirements for shielding of the CO molecule. 

In reality there is unlikely to be a pure dichotomy between the in-situ formation of new molecular gas from a purely ambient atomic medium and the interaction of the shell with a population of fully pre-existing molecular clouds. Nevertheless, this idealization provides a meaningful starting point from which to explore the relationship between supershells and the molecular ISM.

Finally, the evolution of the molecular ISM in supershells links directly to the question of the dominant processes driving star formation within the Galaxy. We round off our discussion by making some brief remarks on star-formation activity within the shell clouds, prior to further dedicated study on this topic.

\subsection{In-Situ Molecular Cloud Formation: Embedded CO Clouds}
\label{insitu}

Supershells enable the rapid production of molecular gas by increasing both the number density and column density of the atomic medium in localized regions. 
The strong dependence of molecule formation rates on density means that the formation of molecules should proceed rapidly in swept-up shells relative to the background, low-density medium. 
Furthermore, the large spatial scales over which supershells operate, and the sheer quantities of material they are able to accumulate over their lifetimes, lead to high column densities within shell walls. This provides shielding from the background UV field, which is essential to the survival of molecular species.


The double filament structure observed in the approaching limb complex of GSH 287+04--17 (figure \ref{fig:287regions} panels \textit{a1} and \textit{a2}; $285.3 < l < 286.4^{\circ}$, $4.2 < b < 6.0^{\circ}$) is a particularly striking example of a feature well explained by the in-situ formation of molecular gas in the swept-up shell. The embedded CO clouds form an integral and coherent part of the curved H{\sc i} wall, with both tracers closely associated in $l$-$b$-$v$ space. High H{\sc i} column densities suggest reservoirs of raw material available for molecule formation, as well as for shielding of the CO molecule. 

Numerical studies suggest that a minimum (1D) extinction of $A_V\sim0.7$ is required for the rapid formation and sustained survival of molecular gas from the atomic medium \citep{bergin04,glover10,wolfire10}. The relation $A_V=N_{\mathrm{HI}}/(1.9\times10^{21}~\mathrm{cm}^{-2})$ \citep[e.g.][]{draine96} may be used in conjunction with the H{\sc i} column density measurements from \S\ref{fitting:coldens} to estimate $A_V$ in the material surrounding the CO clouds. This returns values of $\langle A_V \rangle\sim0.2$  towards lines of sight coincident with embedded CO. However, since $N_{\mathrm{HI}}$ is a lower limit in the case where optical depth is not negligible, this figure is expected to miss a significant fraction of the CNM that dominates the shell walls. Similarly, it cannot account for CO-dark H$_2$, which requires much lower extinctions in order to form, and which must exist in some quantity in envelopes around CO-bright clouds. The contribution from both of these dark fractions may be considerable, and can range from several $10\%$ to several times the CO-bright molecular column density \citep{grenier05,wolfire10}.

Infra-red excess measurements provide a means of evaluating this `dark gas' fraction. Under the assumption of a constant Galactic gas-to-dust ratio, the dust opacity at wavelength $\lambda$, $\tau_\lambda$, is directly proportional to the total hydrogen column density along the line of sight. We make use of reprocessed IRAS 60 $\mu$m and 100 $\mu$m images from the IRIS project \citep{miville05} to evaluate the 100 $\mu$m opacity, $\tau_{100}$. This is then compared with our H{\sc i} and CO datacubes to obtain linear relationships between $\tau_{100}$ and H{\sc i} and CO integrated intensities, and hence to assess the excess column density contribution arising from material not accounted for by either of these two transitions. 

We work with an $8^{\circ}\times6^{\circ}$ region centered on the GSH 287+04--17 approaching limb complex, and follow the prescription of \citet{douglas07} in order to obtain a measurement of the excess opacity $\tau_{100,ex}$. The present analysis differs from their treatment in the following points: 1. We do not explicitly account for the contribution from ionized material, which is generally negligible across the region. 2. We do not apply their `correction factor' that modifies $N_{\mathrm{HI}}$ to include an rough estimate of the contribution from cool gas, since it is explicitly this contribution that we wish to assess.

The results of this analysis are shown in Figure \ref{fig:ir_excess}. Significant excess opacity is observed in the entire section of shell wall, concentrated especially in regions of embedded CO. The highest values of $\tau_{100,ex}$ occur in the vicinity of the small H{\sc ii} region at $(l, b)=(285.9^{\circ}, 4.5^{\circ})$. These are assumed to include a significant contribution from ionized material, and are therefore ignored. However, even outside of this anomalous region, $\tau_{100,ex}$ takes consistently high values, averaging $\sim9\times10^{-5}$ for CO-detected positions. 

The column density implied by this value is estimated by applying the scaling factor relating $\tau_{100}$ and H{\sc i} column density that was obtained during the above analysis. This factor is $\tau_{100}=(4.7\pm0.2)\times10^{-26}~N_{\mathrm{HI}}$, which implies missing hydrogen column densities of $\sim2\times10^{21}$ cm$^{-2}$. This in turn translates to $A_V\sim1$, which combined with the contribution directly traced in H{\sc i}, gives a total extinction of $\sim1.2$. 
Strictly speaking, this figure applies to all material along the line of sight outside of the CO-bright core region, implying a 1D extinction at the surface of the CO cloud of $\sim0.6$. 
However, the parsec-scale resolution of the datasets ensures that this figure underestimates the true peak values of $A_V$ in the region. These results imply that enough material exists in the swept-up wall to permit the existence of a stable core of CO-rich molecular gas. 

The mass of the entire complex is estimated as $\sim2.5\times10^4~M_{\odot}$. Of this, less than half is traced by H{\sc i} and CO, and the remainder is inferred from IR excess, demonstrating the importance of material invisible to these two standard tracers. By assuming that the section of wall possesses a depth somewhere between the minimum and maximum widths it subtends on the sky, an estimate is obtained of the average initial number density required to form the complex via the sweeping up of the pre-shell medium. Taking the center of expansion of the shell as $(l, b)=(287.5^{\circ}, 3.0^{\circ})$ \citep{dawson08a}, results in an initial number density of $\langle n_\mathrm{H}\rangle\sim3-10$ cm$^{-3}$. This is slightly enhanced with respect to the mean density of the pre-shell medium calculated over the entire shell volume from H{\sc i} and CO alone (see \S\ref{totalmass}).
However, it is consistent with a scenario in which initial over-densities within the ambient medium play a role in determining the sites at which efficient molecular cloud formation can occur. 


Finally, it is essential to consider the the formation of molecular gas in supershells in the context of current cloud formation theory. 
Modern theory favors a picture in which clumpy sheets and filaments of CNM are formed through the compression, cooling and fragmentation of WNM in large-scale shocks and colliding flows, of which the case of a supershell expanding into the surrounding medium is but one example \citep[e.g.][]{hennebelle99,audit05,vazquez07,heitsch08}. Densities in these fragments may greatly exceed canonical values for the CNM, as a consequence both of the global ram-pressure and the presence of transient turbulence-driven density enhancements. This increased density leads to fast chemical timescales, and provided that the interaction zone is able to accumulate sufficient material, chemical models suggest that the transition to molecular gas may occur on timescales of $\sim10^{6-7}$ yr \citep{koyama00,bergin04,heitsch08}. In this context it is interesting to note the correlation of embedded CO with complex H{\sc i} velocity profiles (see \S\ref{fitting:deltav}), which may indicate the stirring up of material in interaction zones, or the presence of further local compressive motions within the shell wall, either of which could act to further increase density in localized regions. 

For the conditions appropriate to our objects, the colliding flows model is able to form dense clouds on considerably shorter timescales than alternative methods such as the purely gravitational fragmentation of the swept up shell \citep[e.g.][]{elmegreen01}.
1D models of molecular cloud formation that explicitly include the chemistry of CO and H$_2$ formation suggest that a flow velocity of as little as $\sim10$ km s$^{-1}$ can produce large quantities of CO-rich molecular gas in $\sim10^7$ years for an initial number density of a couple of atoms per cubic centimeter \citep{bergin04}. These figures are consistent with the estimated age and expansion velocity of GSH 287+04--17 
as a whole, and formation timescales should be even shorter for the slightly enhanced initial densities implied for the specific section of shell wall discussed above. 
For GSH 277+00+36, which is both older and more quickly expanding, the conditions for formation are also met easily. Moreover, these formation timescales are expected to become even lower when models are expanded to include realistic, turbulent gas dynamics \citep{glover07, glover10}. 

Magnetic fields are expected to play an important role in this cloud formation process \citep{inoue08, inoue09, heitsch09}. Magnetic support acts to oppose the formation of very dense gas, leading to the additional requirement that the component of $B$ perpendicular to the flow direction must be vanishingly small in order for molecule formation to proceed. This rather stringent requirement may be crucial in pre-selecting the sites where molecular clouds can form, and is pleasingly consistent with the observational reality that supershells contain at best a scattering of molecular clouds. Although at present we have no information on the magnetic fields in either of our shells, this subject should motivate further observational study.

\subsection{Offset CO Clouds: Pre-Existing Molecular Gas}
\label{preexisting}

The approaching limb complex of GSH 287+04--17 also contains several promising candidates for pre-existing molecular clouds. Between $285.5 < l < 288.5^{\circ}$, $5.4 < b < 6.2^{\circ}$ (figure \ref{fig:287regions} panel \textit{a1}), CO-tipped H{\sc i} `fingers' point radially inward toward the shell interior, with the main walls pulled back, arch-like, around them.

In contrast to the embedded clouds described above, these small offset CO-clouds are preferentially associated with very low values of $N_{\mathrm{HI}}$ (see \S\ref{fitting:coldens}). Similarly, there is no significant IR excess coincident with the clouds (see figure \ref{fig:ir_excess}), indicating that there is little or no dark gas undetected in H{\sc i} or CO. Correspondingly, not only is there minimal material for shielding, but there is also no substantial reservoir of raw material available for molecular gas formation. The atomic to molecular mass ratio in CO-detected pixels averages only $\sim 0.1$, and the mean extinction arising from material not traced by CO is estimated from $N_{\mathrm{HI}}$ to be $A_V\sim0.1$. This is an order of magnitude smaller than required to shield CO. We conclude that the ongoing formation of molecular gas in these clouds in their present configuration is not viable. 

A simple check may be performed on a scenario in which the present CO clouds are remnants of molecular gas that pre-dates the shell, by comparing theoretical cloud destruction timescales with the survival times implied by the data. A lower limit to the cloud survival time may be estimated from the shell expansion velocity ($\sim10$ km s$^{-1}$) and cloud-to-main-wall distances ($20-40$ pc), by assuming that the passage of the shell has not accelerated the clouds significantly and ignoring projection effects. This implies that the interaction between the clouds and the wall occurred a minimum of $2-4$ Myr ago. Similar timescales are obtained for the offset clouds in the molecular drip region of GSH 277+00+36 (Figure \ref{fig:277regions}, panels \textit{a1} and \textit{a2}). Here, cloud-to-main-wall distances of $20-30$ pc and an expansion velocity of $\sim20$ km s$^{-1}$ imply survival times of at least $1-1.5$ Myr. 

Dense clouds interacting with an expanding shell will be subject to dynamical disruption from the shell-cloud interaction, followed by thermal evaporation by the hot interior medium in which they become entrained. In addition, the physical stripping and dissipation of cloud material also renders them increasingly vulnerable to the UV dissociation of CO. The observable lifetime of the entity we see as a CO cloud may therefore be shorter than the survival time of the dense material itself.

The dynamical interaction of a dense cloud with a shocked flow is commonly parameterized in terms of the cloud crushing time, $t_{cc}=(r_0/v_i)(n_{cl}/n_i)^{0.5}$, where $r_0$ is the cloud radius, $v_i$ is the velocity of the shock in the ambient medium, and $n_{cl}$ and $n_i$ are the number densities of the cloud and the ambient medium respectively \citep[e.g.][]{klein94}. For the case of a section of shell wall impacting a pre-existing molecular cloud, we adopt the parameters $r_0=2$ pc, $v_i=10$ km s$^{-1}$, $n_{cl}=100$ cm$^{-3}$ and $n_i=10$ cm$^{-3}$, which are appropriate for the shell and clouds in their present state, and correspond to $t_{cc}\approx0.6$ Myr. For the ideal adiabatic case clouds are typically destroyed on timescales of a few to $\sim10$ $t_{cc}$ \citep{klein94,nakamura06}. However, for the present case radiative cooling cannot be ignored over the timescales of the interaction. Radiative cooling tends to inhibit cloud destruction, encouraging the formation of over-dense clumps and filaments, and potentially extending survival times significantly \citep{mellema02,orlando05}. 
This suggests that dynamical disruption of pre-existing molecular clouds will not be so fast as render them unobservable on the relevant timescale of $1-4$ Myr

For a shell thickness of $\sim10$ pc, the cloud will pass through the shell and into the hot interior on timescales of $1$ to several Myr, depending on the drag efficiency. In the interior regime the material flowing past it will be more diffuse and hence less dynamically disruptive, but thermal conduction may become important. The classical evaporation timescale for clouds embedded in fully ionized medium is given by $t_{evap}\sim3.3\times10^{20}~n_{cl}~T_i^{-5/2}r_{cl,\mathrm{pc}}^2$ yr \citep{cowie77}, which equates to $\sim10^8$ yr for the present clouds, assuming $T_i=10^6$ K. Thermal evaporation is therefore unlikely to be important over the lifetime of the shell.

The CO destruction rate by UV may be expressed as $I'_{\mathrm{CO}}G'_0f_{\mathrm{CO}}~e^{-2b_{\mathrm{CO}}A_V}$ s$^{-1}$, where $I'_{\mathrm{CO}}$ is the unshielded CO photodissociation rate in the Draine field, $G'_0$ is the ratio of the incident radiation field to the Draine field, $f_{\mathrm{CO}}$ is the CO self shielding factor, and $b_{\mathrm{CO}}$ is 
the dust extinction coefficient for CO
\citep[see Appendix C of][]{wolfire10}. Values of $I'_{\mathrm{CO}}$ and $f_{\mathrm{CO}}$ are taken from \citet{visser09}, where $f_{\mathrm{CO}}$ is estimated from their table 7, using values of $N_{\mathrm{H}_2}$ from our data and assuming a standard H$_2$ to CO abundance ratio of $10^4$. 
Following \citet{wolfire10}, $b_{\mathrm{CO}}$ is taken as 3.2, $G'_0$ is assumed to be 1, and $A_V$ is estimated as 0.8 based on the values of $N_{\mathrm{H}_2}$ and $N_{\mathrm{HI}}$ at the spatial centers of the CO clouds. This gives an estimated lifetime of $\tau_{\mathrm{CO}} 
\sim 5$ Myr for CO at the center of the clouds in their current form, implying that at least some molecular material should remain observable for several million years.
In contrast, the survival time at the surface of the CO clouds where $A_V\approx0.1$ is as little as $\sim 10^4$ yr.
This is consistent with a scenario in which the dynamical stripping of material from around an originally stable CO cloud renders it increasingly vulnerable to UV dissociation. 

While all of the timescales estimated above are necessarily rough, they nevertheless illustrate the plausibility of a scenario in which many of the offset clouds we observe in our two shells are remnants of pre-existing molecular material. However, it must be noted that not all offset clouds in the present sample are so neatly explained by this simple scenario. Particularly those at very high latitudes, for which it is difficult to postulate the a-priori existence of a large quantity of pre-existing molecular gas. A more realistic description of supershell evolution might contain `hybrid' scenarios, in which clumpy shell walls containing newly-formed CO clouds are carved by later episodes of energy input into the configurations now observed. 

Finally, we note that we have not commented on the possible role of Rayleigh-Taylor instabilities in creating shell structure, which has been suggested as a possible means of forming the H{\sc i} `drips' observed in GSH 277+00+36 \citep{mcclure03}. The presence of pre-existing dense structure in the inhomogeneous ambient medium is likely to affect the development of the instability, modifying the properties of the RT fingers that form \citep[e.g.][]{jun96}. However, a full consideration of the interplay between instabilities in the shell wall and pre-existing dense clouds is beyond the scope of this paper.

\subsection{Remarks on Star Formation}
\label{stars}

Since molecular clouds are the sites of star formation, the creation and destruction of molecular material in supershells is of potentially critical importance to Galactic star formation rates.

\citet{fukui99} compared the molecular cloud population of GSH 287+04--17 with young stellar object (YSO) candidates extracted from the IRAS point source catalogue, finding a number of convincing correlations that strongly argued for ongoing star formation in the shell clouds. Figure \ref{fig:287iraspsc} reproduces this source list for GSH 287+04--17, while Figure \ref{fig:277iraspsc} extends the analysis to GSH 277+00+36. The present high-resolution comparison of CO and H{\sc i} allows us to place these potential star forming regions in their proper context within the ISM of the shell walls. 

The molecular clouds of the GSH 287+04--17 approaching limb complex show excellent correlation with a number of IRAS sources. Particularly striking is the presence of YSO candidates in all three of the small offset CO clouds discussed in \S\ref{preexisting}. As promising examples of molecular material pre-dating the shell, this raises the possibility that star formation is being triggered in these clouds as a direct result of their interaction with the shell. Indeed, in shock-cloud interactions where radiative cooling is important, compression and cooling may lead to localized gravitational collapse even as much of the mass of the cloud is stripped away \citep{foster96,fragile04,vanloo07}. Similarly, the larger of the two offset clouds in the molecular drip region of GSH 277+00+36 is also associated with a YSO candidate.

Perhaps more importantly, the embedded molecular cloud in the approaching limb complex of GSH 287+06-17 discussed in \S\ref{insitu} contains an active massive star forming region at $(l, b)=(285.9,4.5)$. This region contains a small optical H{\sc ii} region (see Figure \ref{fig:ir_excess}), two luminous IRAS YSOs, and the recently confirmed 2MASS cluster DBSB 49 \citep{dutra03}. The most massive member of this small open cluster is of spectral type B0V, and it has an estimated age of $2.1\pm0.3$ Myr \citep{soares08}, placing the onset of star formation relatively recently in the shell's history. Given this cloud's proposed origin as an object formed in-situ in the shell wall, this suggests the triggered formation of both molecular gas \textit{and} stars in ISM flows on total timescales of $\sim10^7$ yr. This is highly consistent with a picture in which the onset of star formation in locally self-gravitating clumps occurs rapidly after sufficiently dense conditions are reached for the creation of the parent molecular cloud \citep{hartmann01,vazquez07}. 

Embedded molecular material in GSH 277+00+36 shows similarly strong associations with luminous IRAS sources, as well as other clear indicators of star formation activity. In particular, the giant H{\sc ii} region RCW42 ($l$=$274.0^{\circ}$, $b$=$-1.1^{\circ}$) is located within the prominent embedded CO complex discussed in \S\ref{results:277}, attesting to active massive star formation in the shell walls. This region also contains the deeply embedded 2MASS stellar cluster DBSB 38 \citep{dutra03}, and several other luminous YSO candidates associated with the wider CO cloud complex. In addition, several other close associations between IRAS sources and CO clouds in both shells are shown in Figures \ref{fig:287iraspsc} and \ref{fig:277iraspsc}. 

Taken together, these results attest to the ongoing formation of stars throughout the molecular material in both GSH 287+04--17 and GSH 277+00+36. We suggest that future studies might build on these preliminary findings in order to better quantify the global star formation rates in supershell-associated molecular clouds, and explore their wider role in the Galactic ecosystem.

\section{Conclusions}
\label{conclusions}

We have presented parsec-scale resolution observations of H{\sc i} and $^{12}$CO(J=1--0) in two Galactic supershells, GSH 287+04--17 and GSH 277+00+36, with the aim of investigating the role played by supershells in the evolution of the molecular ISM. The main findings may be summarized as follows.

1. Both shells contain large quantities of associated molecular gas in the form of discrete co-moving CO clouds distributed throughout the atomic shell walls. These high resolution observations reveal rich substructure in both tracers. Molecular gas is seen elongated along the inner edges of atomic shell walls, embedded within H{\sc i} filaments and clouds, or taking the form of small CO clouds at the tips of tapering `fingers' of H{\sc i}. We note a similarity in features observed in both objects, despite differences in location and evolutionary stage. 

2. The atomic shell walls are dominated by cold gas, showing narrow linewidths reaching as low as $\Delta v\sim2$ km s$^{-1}$. Mean temperatures and densities are estimated to be roughly $T_k\sim100$ K and $n_0\sim10$ cm$^{-3}$ in regions where the spectral signature from the shells is well determined. 

3. An enhanced level of molecularization is observed over the volumes of both GSH 287+04--17 and GSH 277+00+36, providing the first direct observational evidence of increased molecular cloud production due to the influence of supershells. Our results imply that the amount of molecular matter in the volumes affected by these shells is enhanced by $\sim3$ times with respect to neighboring regions. If this is found to hold true on Galactic scales, it has powerful implications for our understanding of the role played by stellar feedback on the evolution of the molecular phase. Already, the confirmation of this phenomenon in two shells at quite different locations and evolutionary stages is very compelling, and we suggest that more followup work should be undertaken to repeat this analysis for other objects. 

4. CO clouds embedded in the main atomic shell walls provide excellent candidates for the in-situ formation of molecular gas from the swept up medium. This scenario has been explored in detail for archetypal examples of embedded molecular clouds, demonstrating that the formation timescales implied by theory are consistent with shell ages and number densities, and that the requirements for the shielding of the CO molecule are met. We thus confirm on a local scale the viability of the triggered formation of molecular clouds due to the influence of the shells. 

5. Small offset CO clouds located at the tips of tapering `fingers' of H{\sc i} may be the remnants of molecular gas present in the ISM prior to the formation of the shells. We have demonstrated the plausibility of this scenario for archetypal examples of such offset clouds, by comparing estimates of cloud destruction times with the survival times implied by the data. 

6. A preliminary examination of YSO candidates in the shell regions confirms that active star formation is occurring in the molecular clouds in both GSH 287+00-17 and GSH 277+00+36. This includes massive star forming regions in embedded molecular cloud complexes, as well as
less luminous YSO candidates associated with offset CO clouds. \\

We wish to thank the anonymous referee for comments that led to the improvement of this manuscript. We also thank Shu-ichiro Inutsuka and Tsuyoshi Inoue, whose helpful discussions have contributed to this work. We gratefully acknowledge the past staff and students of Nagoya University who made the CO observations utilized in this paper. The NANTEN project was based on a mutual agreement between Nagoya University and the Carnegie Institute of Washington, and its operation was made possible thanks to contributions from many companies and members of the Japanese public. The Australia Telescope Compact Array and Parkes Telescope are part of the Australia Telescope which is funded by the Commonwealth of Australia for operation as a National Facility managed by CSIRO. This research has made use of the NASA/IPAC Infrared Science Archive, which is operated by the Jet Propulsion Laboratory, California Institute of Technology, under contract with the National Aeronautics and Space Administration. We also gratefully acknowledge the Southern H-Alpha Sky Survey Atlas (SHASSA), which is supported by the National Science Foundation. 

\bibliographystyle{apj}
\bibliography{newbibliography}

\begin{figure}
\epsscale{0.8}
\plotone{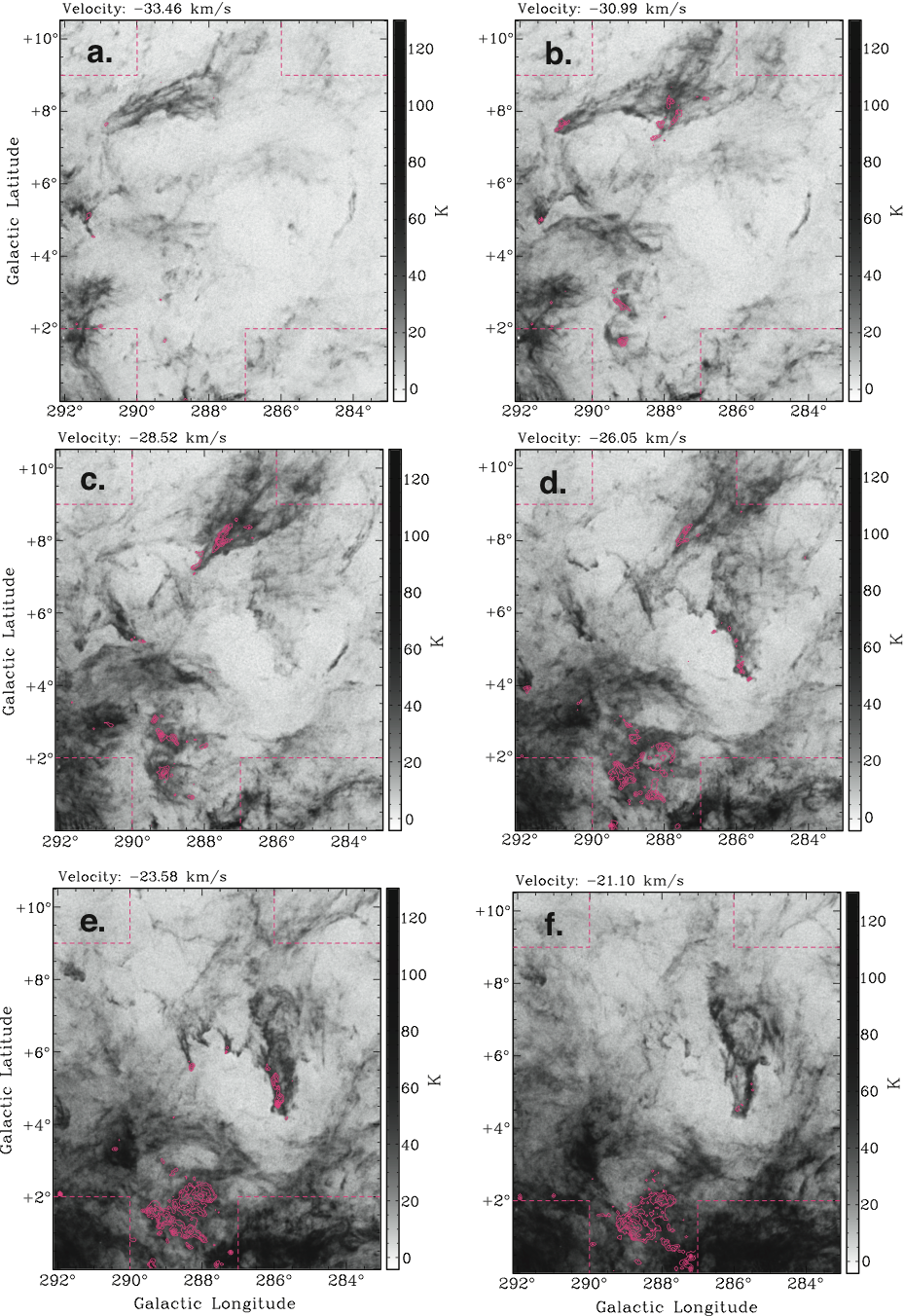}
\caption{Channel maps of the GSH 287+04--17 region. Each panel shows mean brightness temperature over a velocity interval of 2.47 km s$^{-1}$. Greyscale images are H{\sc i} data. Pink contours show $^{12}$CO(J=1--0), beginning at a starting level of 0.7 K and incremented in steps of 1.0 K. Dotted lines mark the boundary of the area observed in $^{12}$CO(J=1--0). The white dashed line in panel g marks the approximate outline of the main body of the shell.}
\label{fig:287chs}
\end{figure}

\addtocounter{figure}{-1}

\begin{figure}
\epsscale{0.8}
\plotone{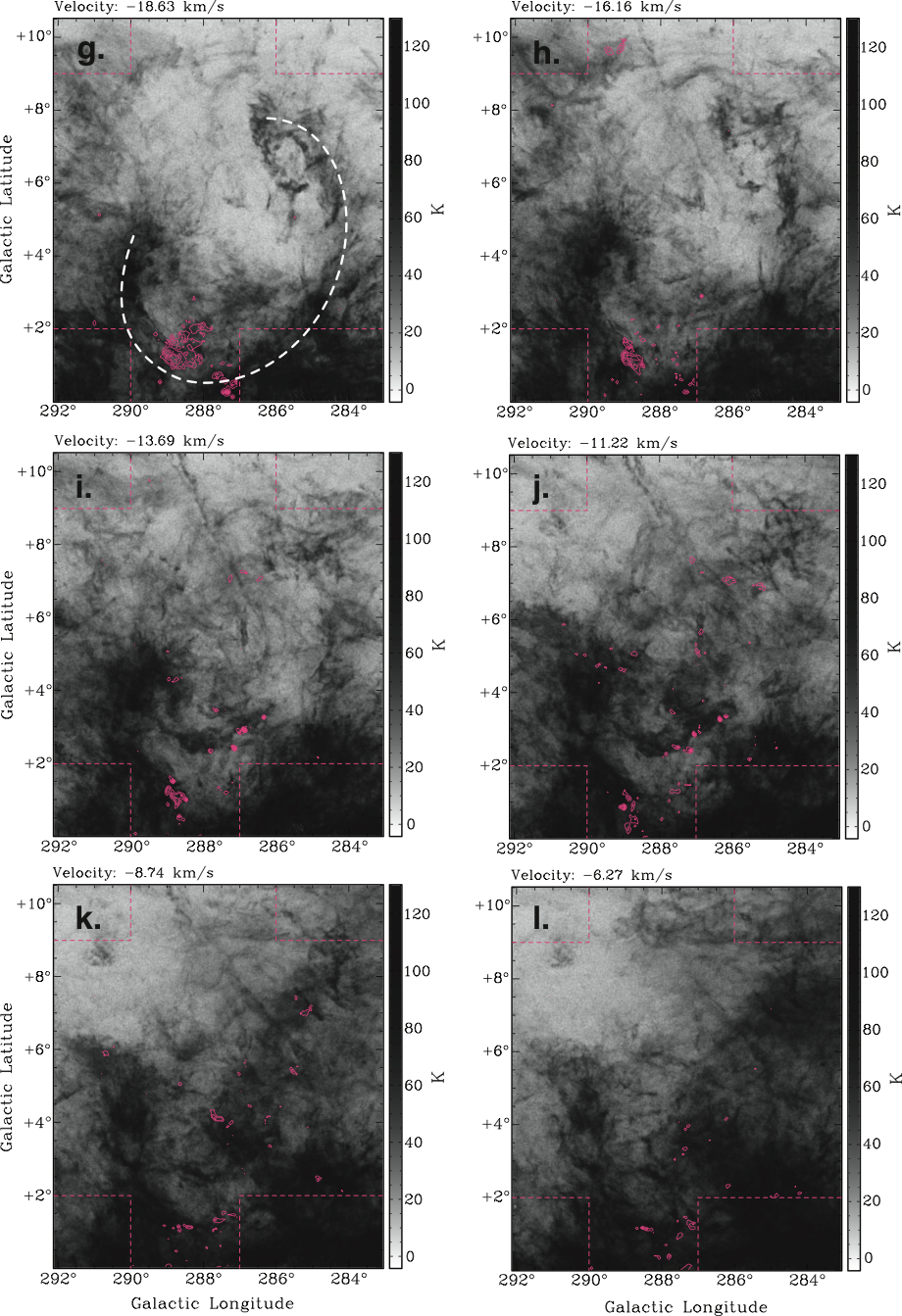}
\caption{cont.}
\end{figure}

\begin{figure}
\epsscale{0.8}
\plotone{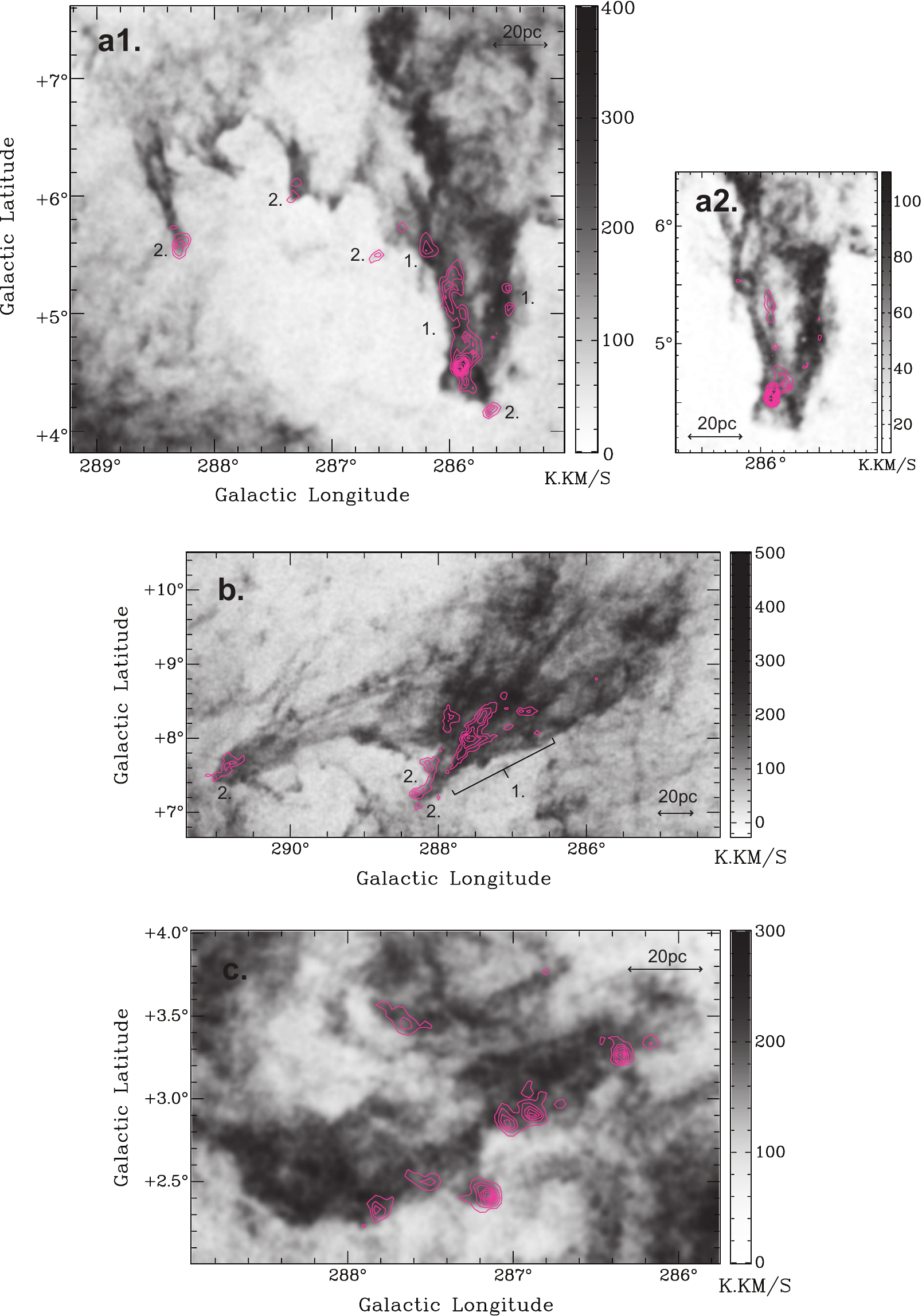}
\caption{Integrated intensity images showing subregions of interest in GSH 287+04--17. Greyscale images are H{\sc i} and pink contours are $^{12}$CO(J=1--0). The velocity integration ranges and contour levels are as follows: (a1) $-26.5 < v_{lsr} < -19.9$ km s$^{-1}$, 1.5+3.0 K km s$^{-1}$; (a2) $-23.2 < v_{lsr} < -21.5$ km s$^{-1}$, 1.3+1.0 K km s$^{-1}$; (b) $-33.0 < v_{lsr} < -25.6$ km s$^{-1}$, 1.5+5.0 K km s$^{-1}$; (c) $-14.1 < v_{lsr} < -10.8$ km s$^{-1}$, 1.5+3.0 K km s$^{-1}$. Panels (a1) and (b) show the regions referred to in the text as the `approaching limb complex' and  `high latitude complex', respectively. CO clouds labelled 1 and 2 indicate those specifically referred to as `embedded' and `offset' in the text.}
\label{fig:287regions}
\end{figure}

\begin{figure}
\epsscale{0.8}
\plotone{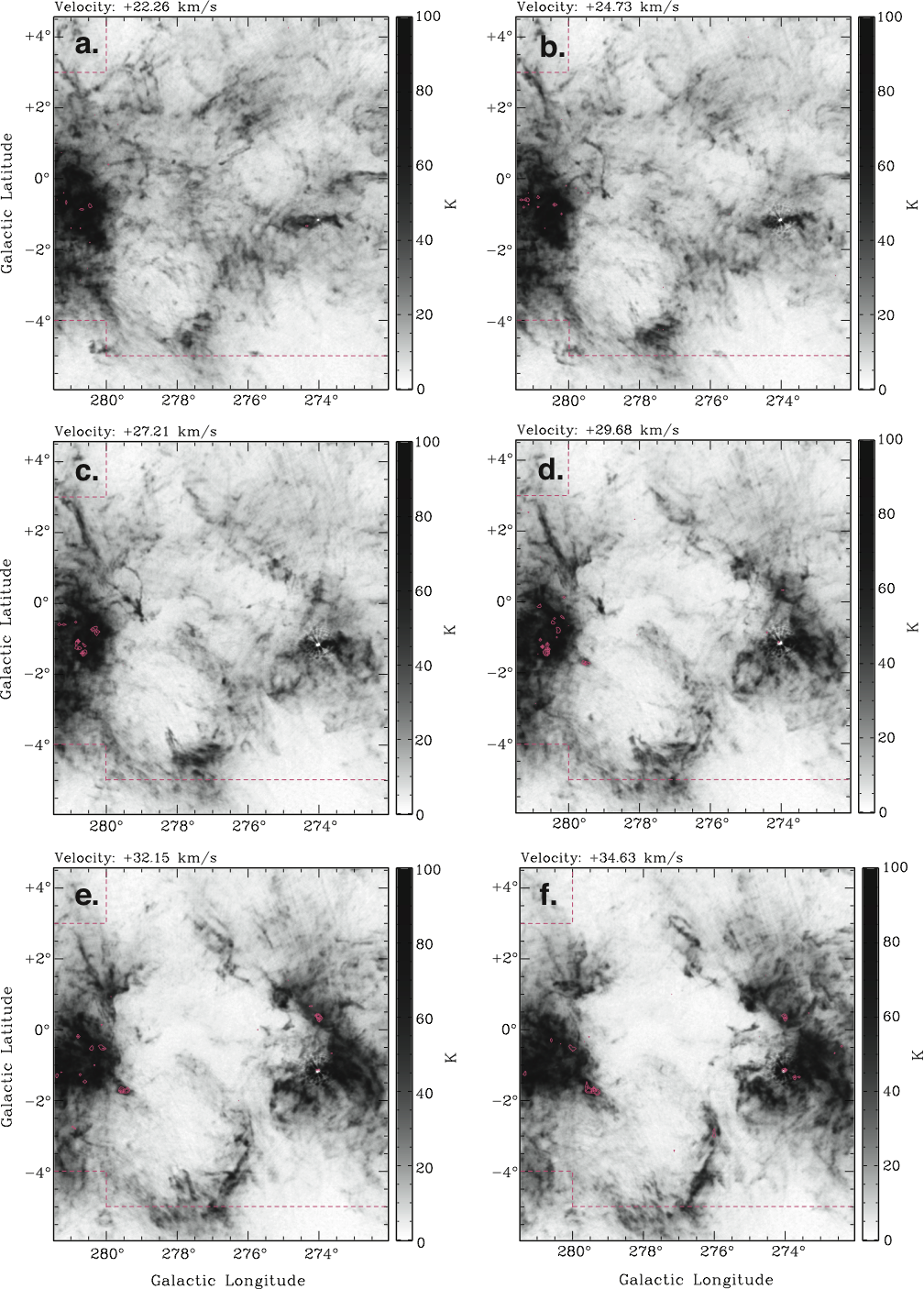}
\caption{Channel maps of the GSH 277+00+36 region. Each panel shows mean brightness temperature over a velocity interval of 2.47 km s$^{-1}$. Greyscale images are H{\sc i} data. Pink contours show $^{12}$CO(J=1--0), beginning at a starting level of 0.8 K and incremented in steps of 1.0 K. Dotted lines mark the boundary of the area observed in $^{12}$CO(J=1--0).}
\label{fig:277chs}
\end{figure}

\addtocounter{figure}{-1}

\begin{figure}
\epsscale{0.8}
\plotone{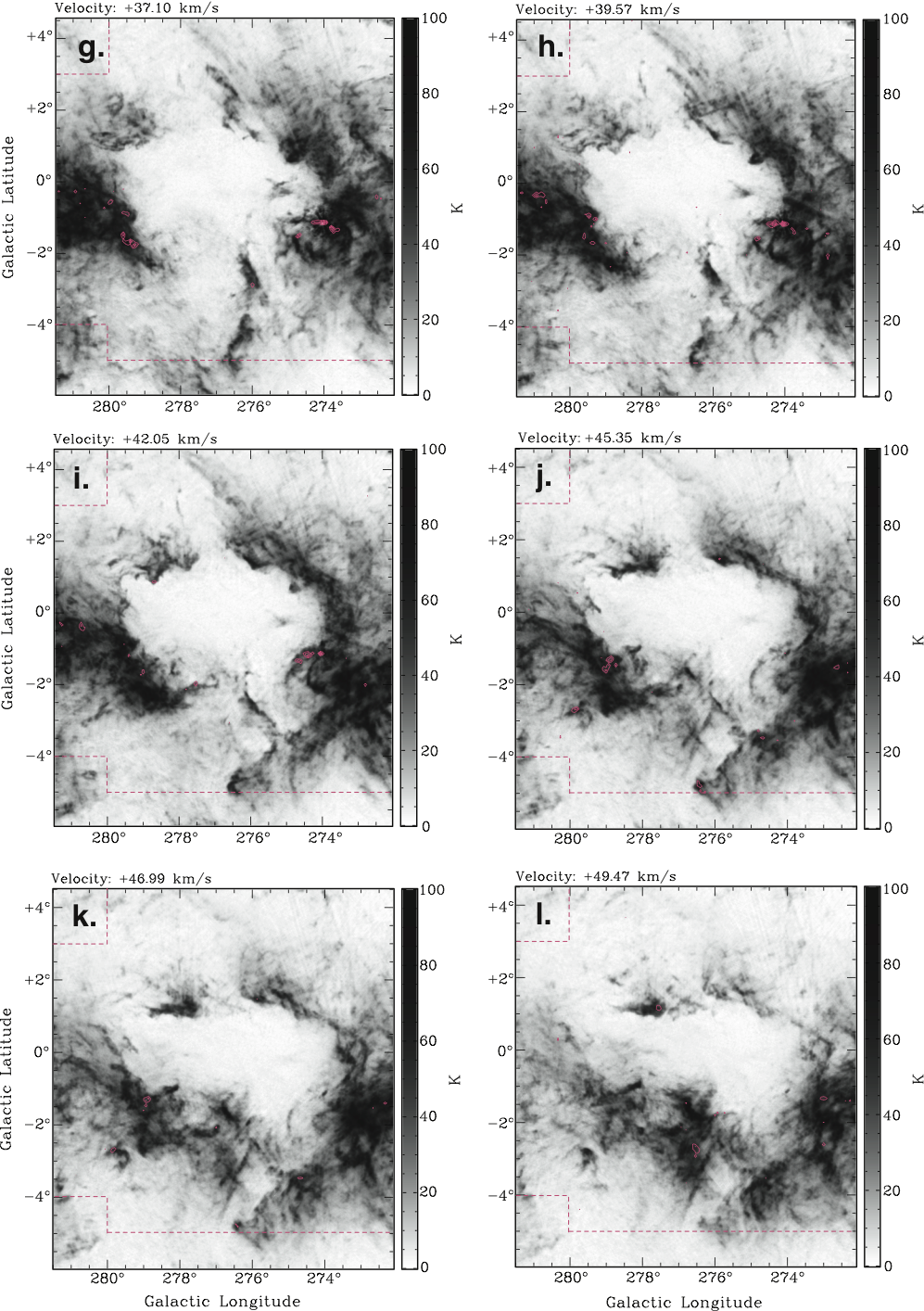}
\caption{cont.}
\end{figure}

\addtocounter{figure}{-1}

\begin{figure}
\epsscale{0.8}
\plotone{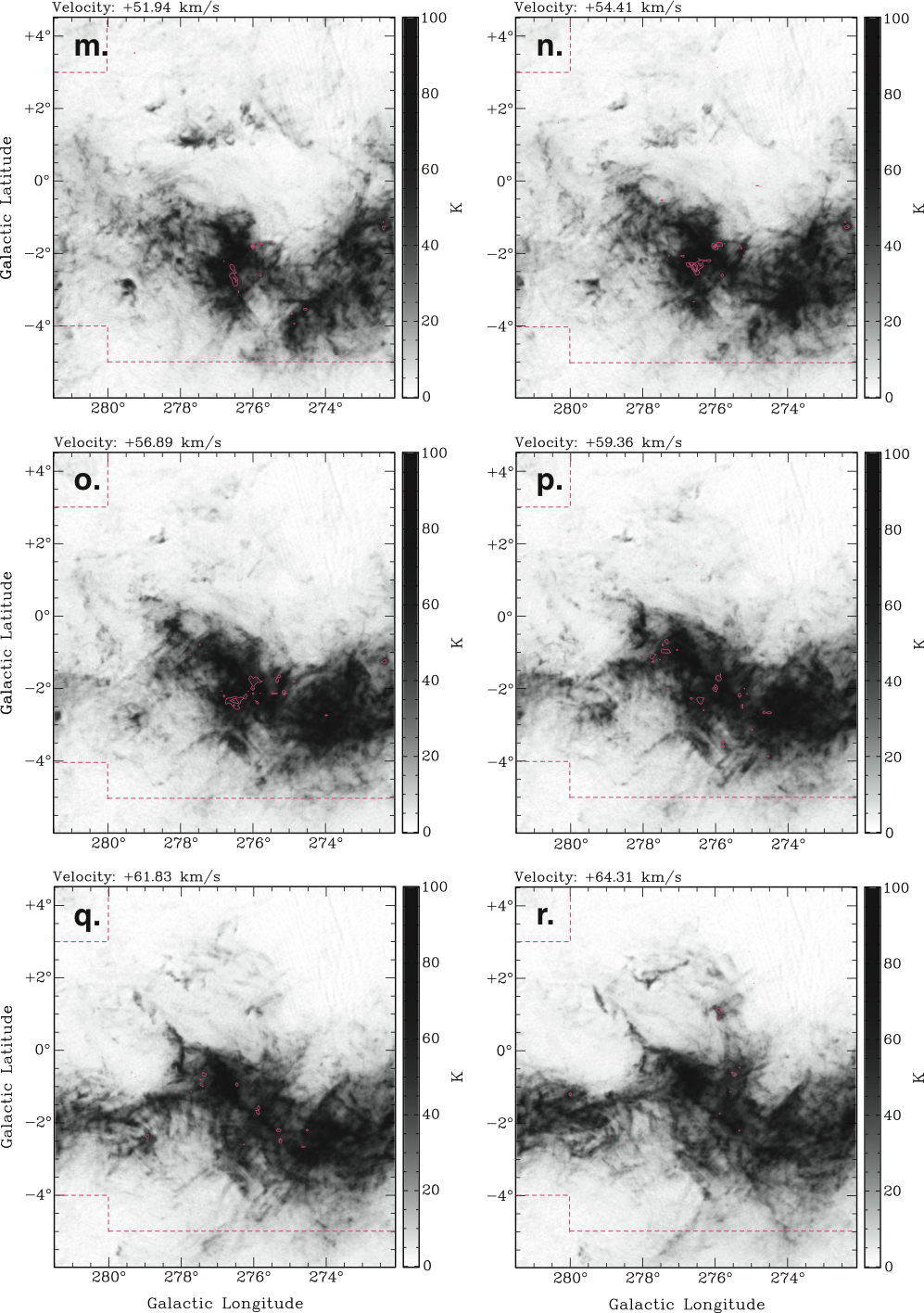}
\caption{cont.}
\end{figure}

\begin{figure}
\epsscale{1.0}
\plotone{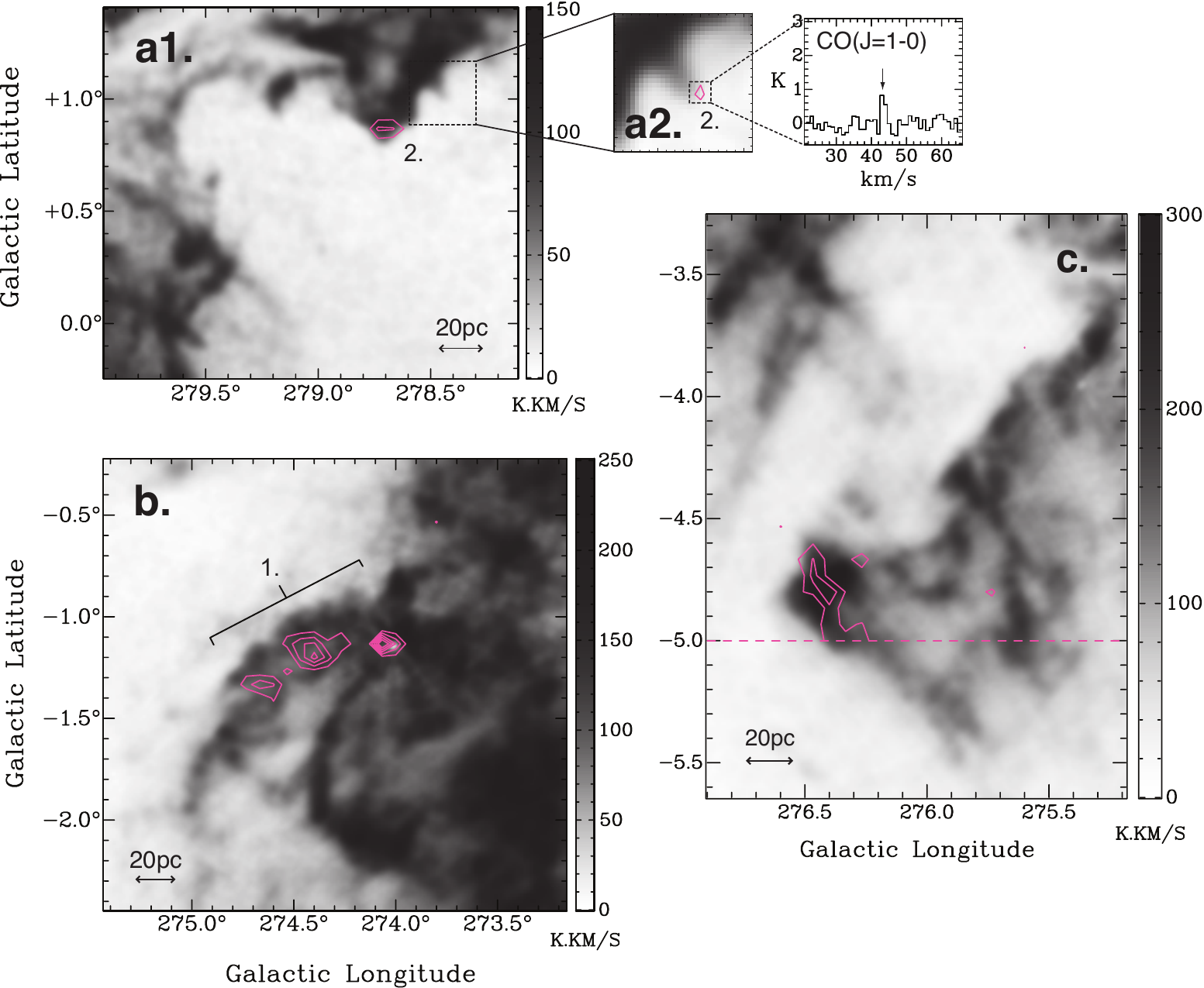}
\caption{Integrated intensity images showing subregions of interest in GSH 277+00+36. Greyscale images are H{\sc i} and pink contours are $^{12}$CO(J=1--0). The velocity integration ranges and contour levels are as follows: (a1) $40.8 < v_{lsr} < 43.3$ km s$^{-1}$, 1.5+1.0 K km s$^{-1}$ (a2) $42.5 < v_{lsr} < 44.1$ km s$^{-1}$ 1.2+1.0 K km s$^{-1}$; (b) $40.8 < v_{lsr} < 43.3$ km s$^{-1}$, 1.5+2.0 K km s$^{-1}$; (c) $43.3 < v_{lsr} < 48.2$ km s$^{-1}$, 1.5+1.0 K km s$^{-1}$. Panel (a1) shows the region referred to in the text as the `molecular drip region'. The spectrum shows CO emission from the single-pixel cloud in panel a2. CO clouds labelled 1 and 2 indicate those specifically referred to as `embedded' and `offset', in the text.}
\label{fig:277regions}
\end{figure}

\begin{figure}
\plotone{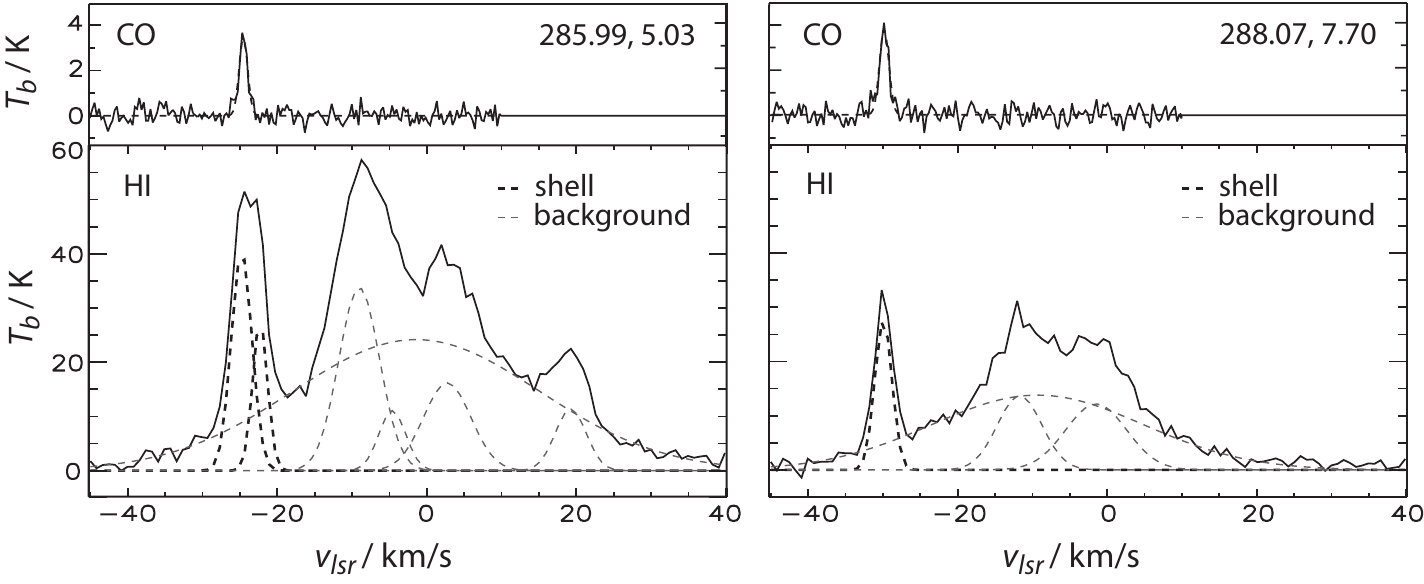}
\caption{Examples of Gaussian decomposition of H{\sc i} and $^{12}$CO(J=1--0) spectra in the GSH 287+04--17 region. Solid black lines show observed spectra. Dashed black lines show Gaussian components fitted to signal emission arising from the shell walls. Dashed grey lines show components fitted to non-related background emission.}
\label{fig:examplespecs}
\end{figure}

\begin{figure}
\plotone{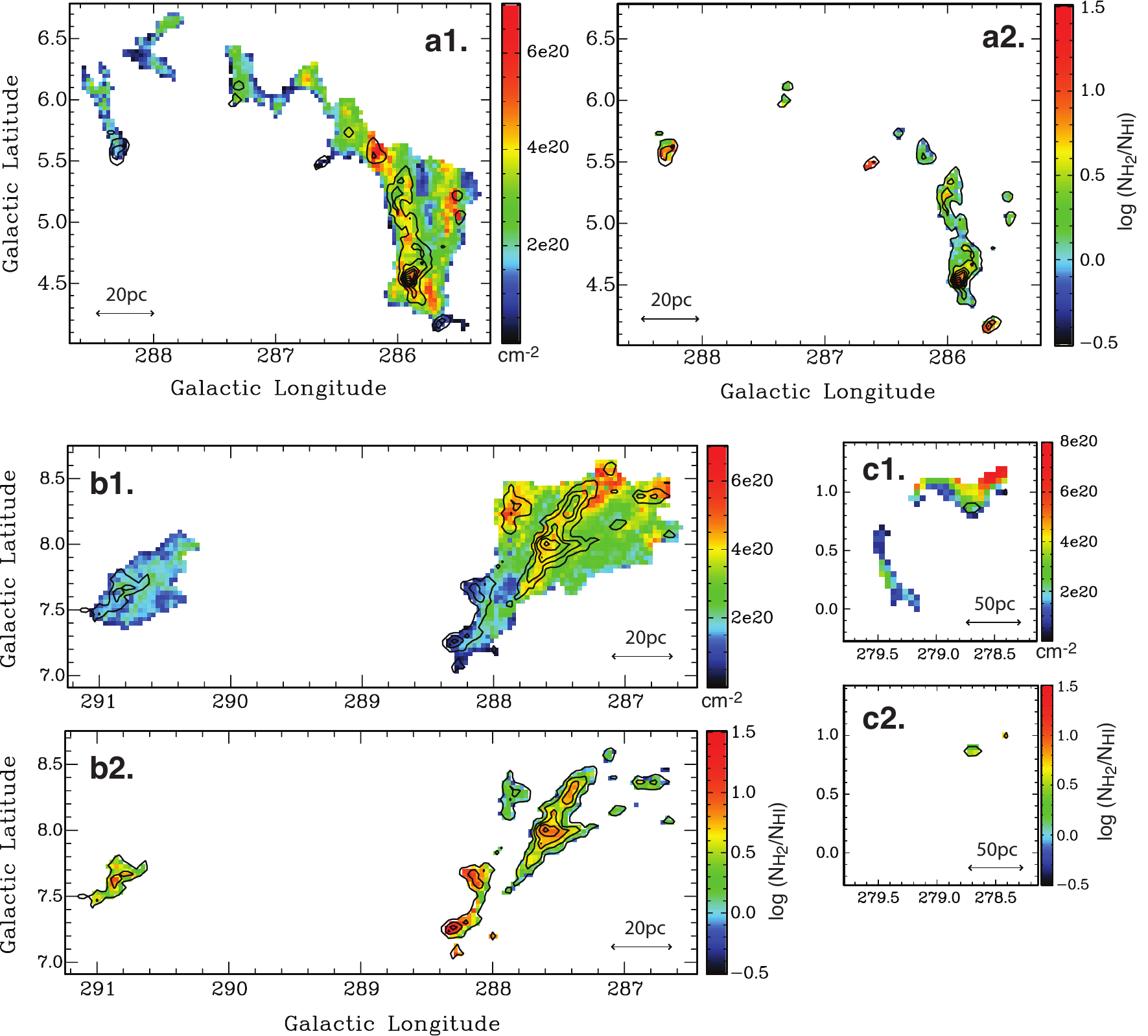}
\caption{$N_{\mathrm{HI}}$ (panels labelled 1) and $N_{\mathrm{H2}}/N_{\mathrm{HI}}$ (panels labelled 2) derived from Gaussian fits to spectra in selected regions of the two supershells. Panels a and b show the approaching limb and high latitude complexes of GSH 287+04--17. Panel c shows the molecular drip region of GSH 277+00+36. Black contours are $N_{\mathrm{H2}}$ beginning at a starting level of $2.4\times10^{20}$ cm$^{-2}$ and incremented in steps of $8.0\times10^{20}$ cm$^{-2}$.}
\label{fig:nhimaps}
\end{figure}

\begin{figure}
\epsscale{0.5}
\plotone{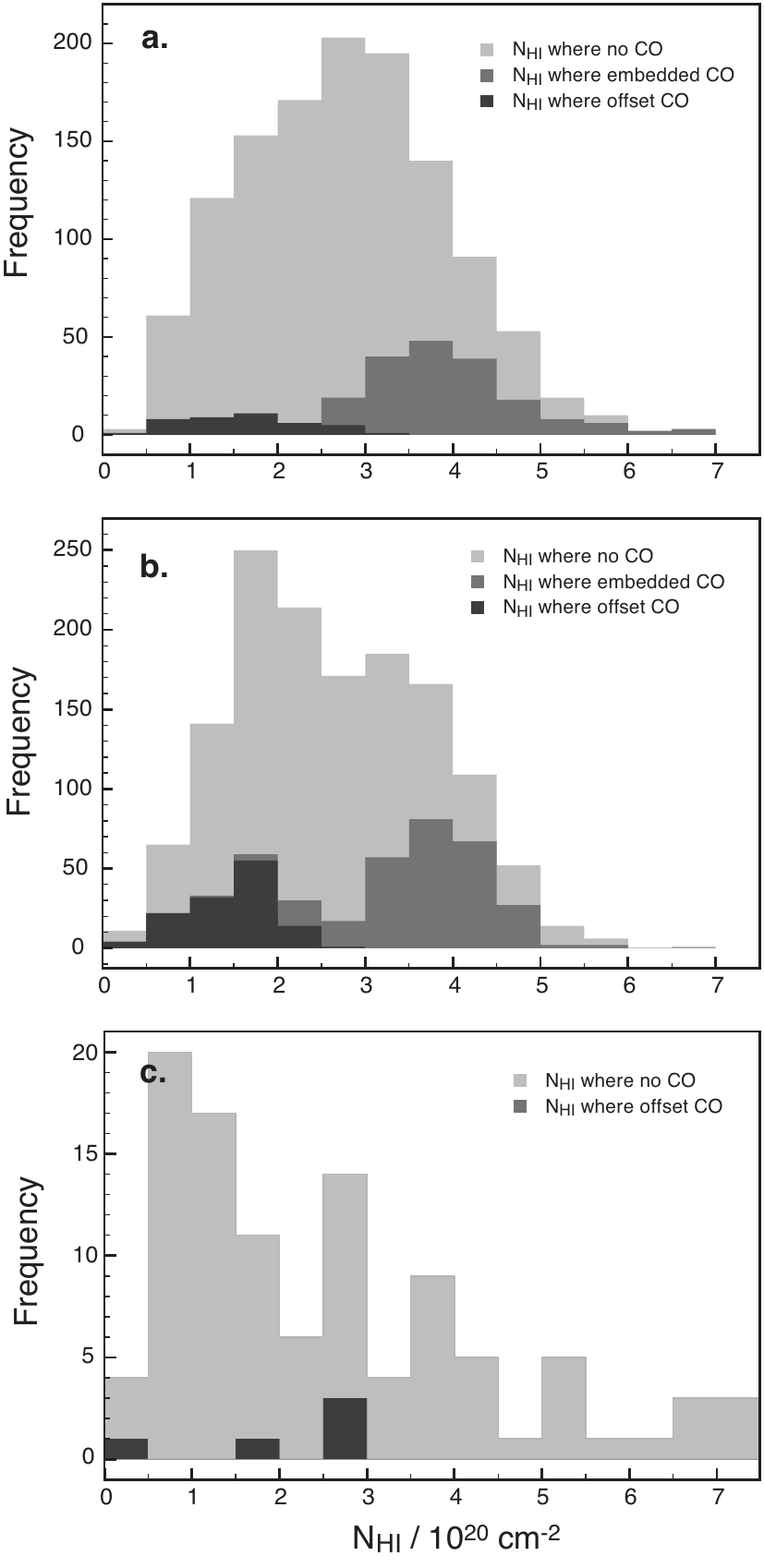}
\caption{Histograms of $N_{\mathrm{HI}}$ derived from Gaussian fits to spectra in selected regions of the two supershells. The charts are color-coded according to the presence or absence of $^{12}$CO(J=1--0) emission at a given position. Panels a and b show the approaching limb and high latitude complexes of GSH 287+04--17. Panel c shows the molecular drip region of GSH 277+00+36.}
\label{fig:nhihist}
\end{figure}

\begin{figure}
\epsscale{1.0}
\plotone{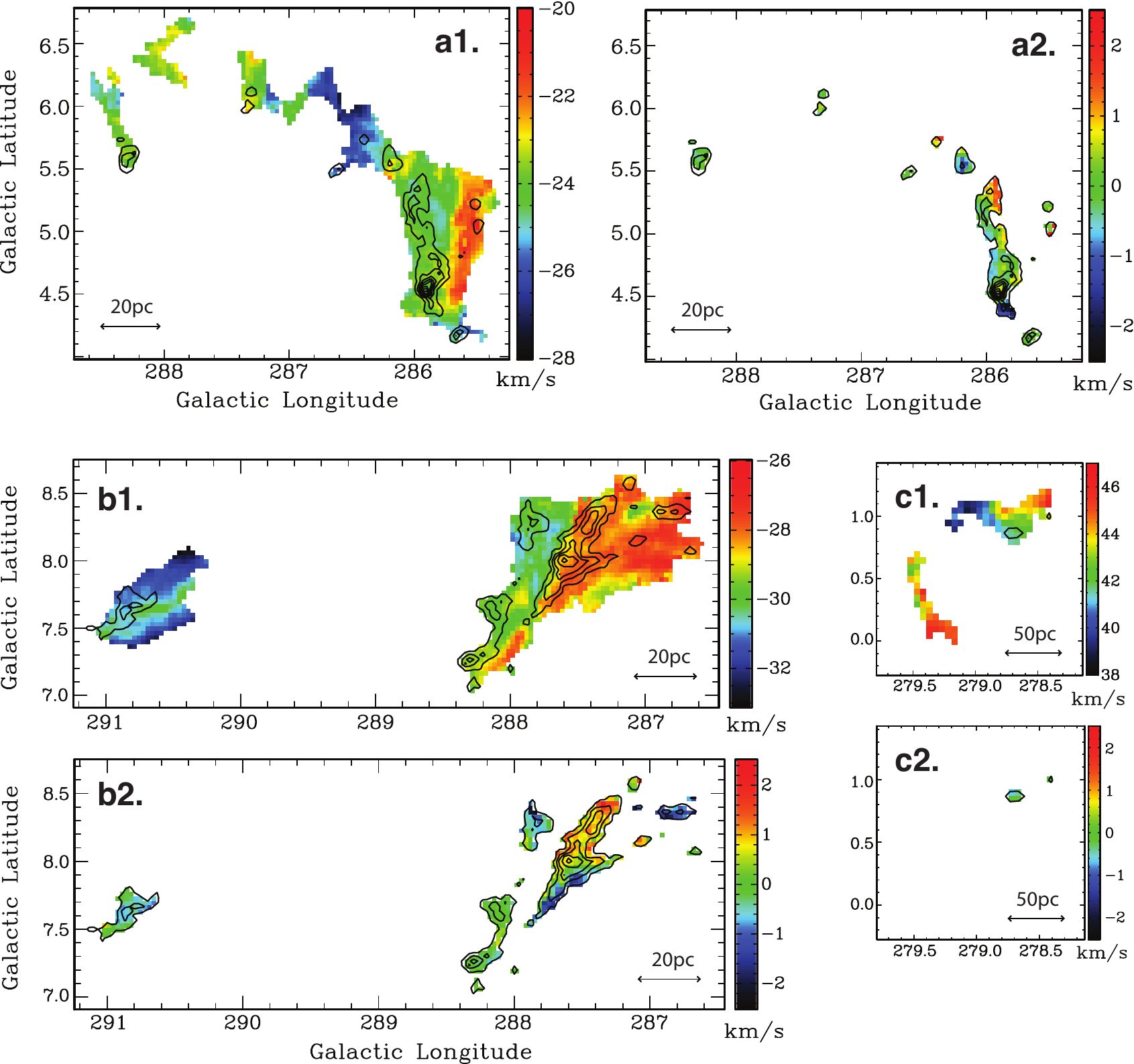}
\caption{H{\sc i} intensity-weighted mean velocity, $v_{0,\mathrm{HI}}$ (panels labelled 1), and the velocity offset between H{\sc i} and $^{12}$CO(J=1--0), $(v_{0,\mathrm{CO}}-v_{0,\mathrm{HI}})$ (panels labelled 2), derived from Gaussian fits to spectra in selected regions of the two supershells. Maps a and b show the approaching limb and high latitude complexes of GSH 287+04--17. Map c shows the molecular drip region of GSH 277+00+36. Black contours are $N_{\mathrm{H2}}$ beginning at a starting level of $2.4\times10^{20}$ cm$^{-2}$ and incremented in steps of $8.0\times10^{20}$ cm$^{-2}$.}
\label{fig:v0maps}
\end{figure}

\begin{figure}
\epsscale{1.0}
\plotone{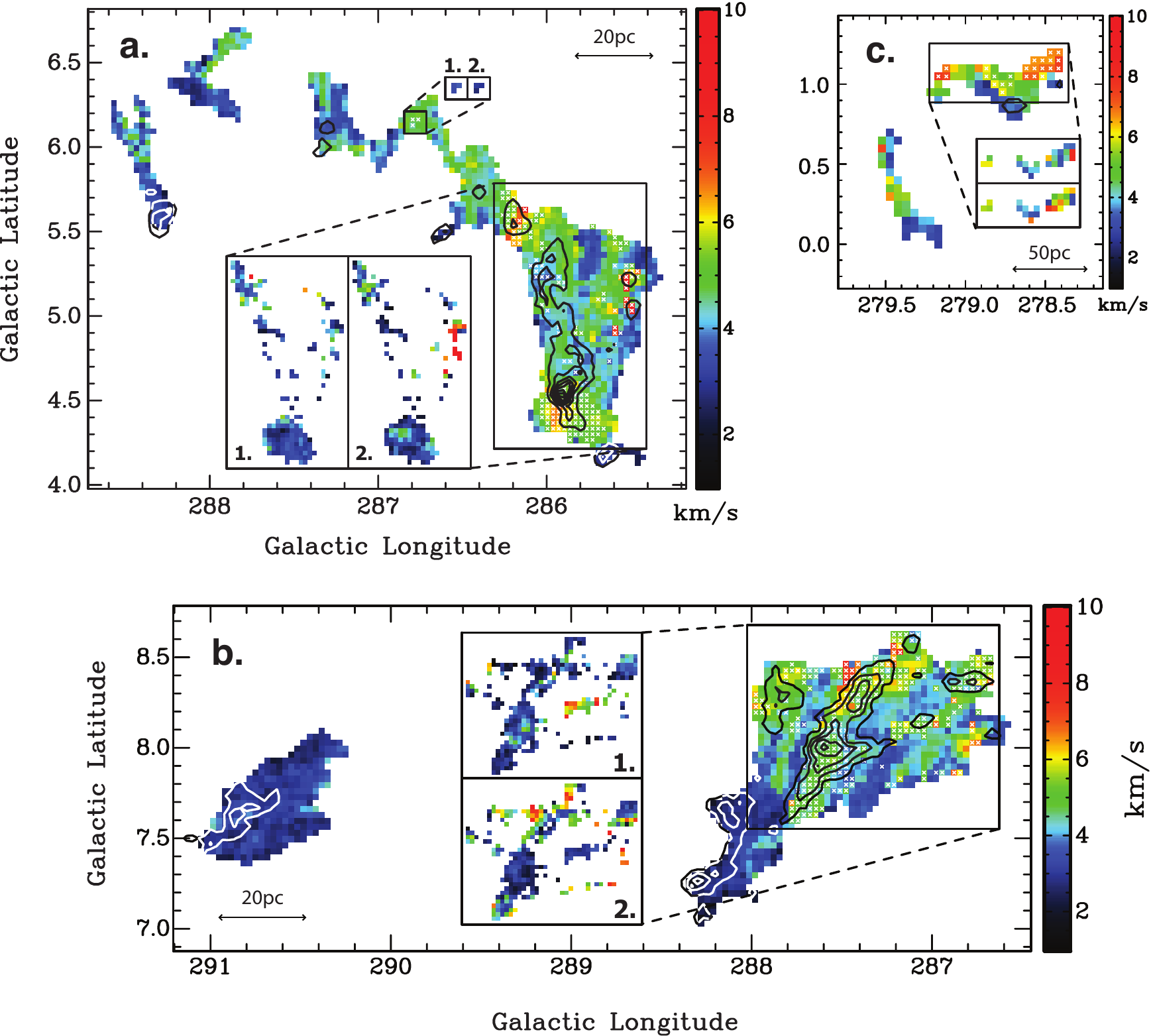}
\caption{H{\sc i} velocity dispersion, derived from Gaussian fits to spectra in selected regions of the two supershells. Panels a and b show the approaching limb and high latitude complexes of GSH 287+04--17. Panel c shows the molecular drip region of GSH 277+00+36. The main maps show the quantity $\Delta v_{\mathrm{HI},tot}=\sqrt{8ln(2)}~\sigma_{v,tot}$, where $\sigma_{v,tot}$ is the intensity-weighted standard deviation of all shell wall emission at a given position. For positions where the shell emission was best fitted by two Gaussian components, inset panels show  the velocity dispersions, $\Delta v_{\mathrm{HI},1}$ and $\Delta v_{\mathrm{HI},2}$ of these components separately. These locations are also marked with white crosses. Contours are $N_{\mathrm{H2}}$, starting at a level of $2.4\times10^{20}$ cm$^{-2}$ and incremented in steps of $8.0\times10^{20}$ cm$^{-2}$.}
\label{fig:lwmaps}
\end{figure}

\begin{figure}
\epsscale{0.5}
\plotone{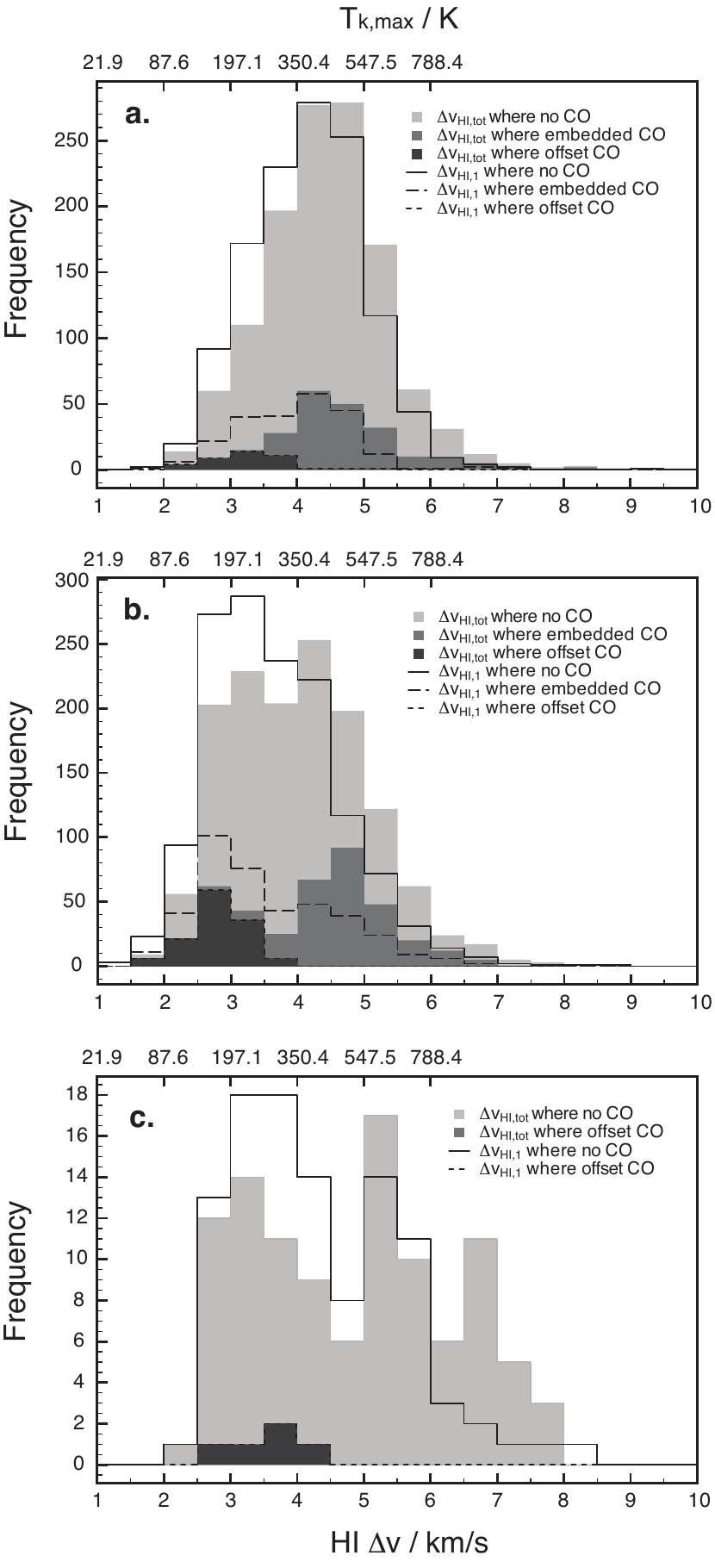}
\caption{Histograms of $\Delta v_{\mathrm{HI}}$ derived from Gaussian fits to spectra in selected regions of the two supershells. Grey shaded charts show $\Delta v_{\mathrm{HI},tot}$, color-coded according to the presence or absence of $^{12}$CO(J=1--0) emission. Solid, dashed and dotted lines delineate the equivalent regions for  $\Delta v_{\mathrm{HI},1}$. The top axis shows the upper limit on kinetic temperature, given by $T_{k,max}=21.9~\Delta v^2$. Panels a and b show the approaching limb and high latitude complexes of GSH 287+04--17. Panel c shows the molecular drip region of GSH 277+00+36.} 
\label{fig:lwhist}
\end{figure}

\begin{figure}
\epsscale{1.0}
\plotone{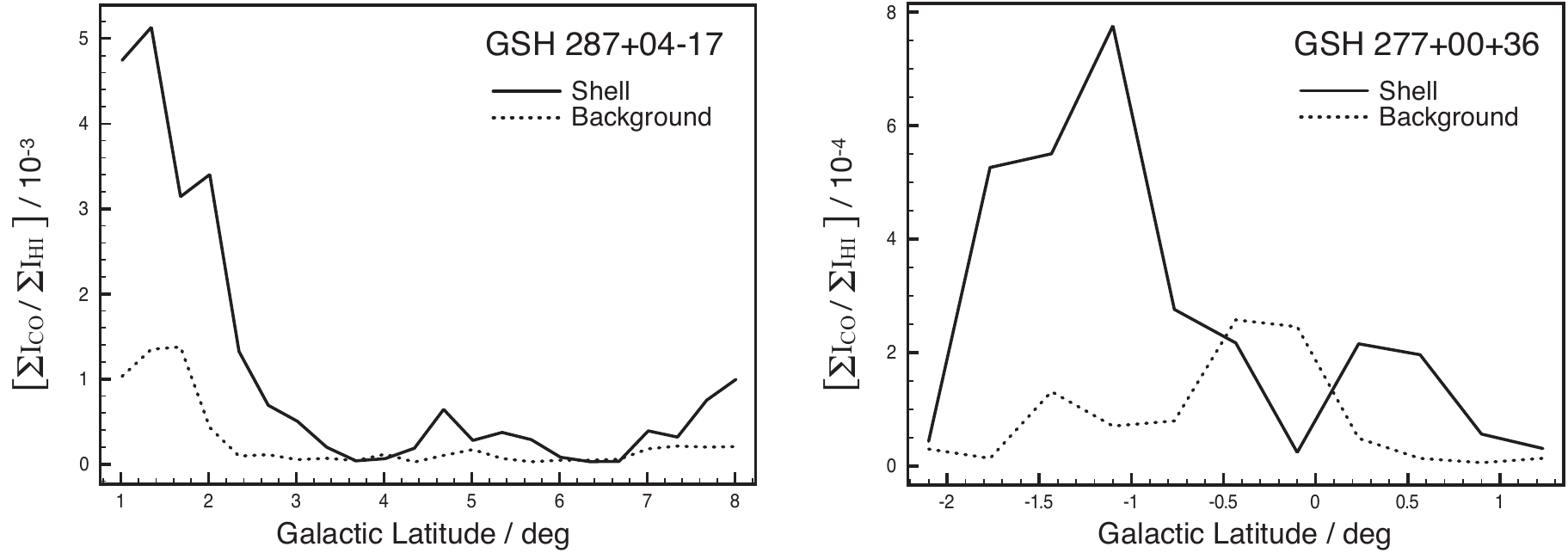}
\caption{Variation of $[\sum I_{\mathrm{CO}}/\sum I_{\mathrm{HI}}]_{sh}$ and $[\sum I_{\mathrm{CO}}/\sum I_{\mathrm{HI}}]_{bg}$ with Galactic latitude. The spatio-velocity ranges of the sub-cubes used in this analysis are as follows: GSH 287+04--17: $284.0 < l < 293.0^{\circ}$, $0.83 < b < 8.17^{\circ}$, $-40.0 < v_{lsr} < -10.0$ km s$^{-1}$;  GSH 277+00+36: $270.0 < l < 282^{\circ}$, $-2.27 < b < 1.47^{\circ}$, $20.0 < v_{lsr} < 50.0$ km s$^{-1}$. Plotted $[\sum I_{\mathrm{CO}}/\sum I_{\mathrm{HI}}]_{sh}$ and $[\sum I_{\mathrm{CO}}/\sum I_{\mathrm{HI}}]_{bg}$ data points are calculated by summing the total CO and H{\sc i} integrated intensities over the defined shell and background regions within slices each spanning $0.33^{\circ}$ in latitude.}

\label{fig:coenhancement}
\end{figure}

\begin{figure}
\epsscale{0.6}
\plotone{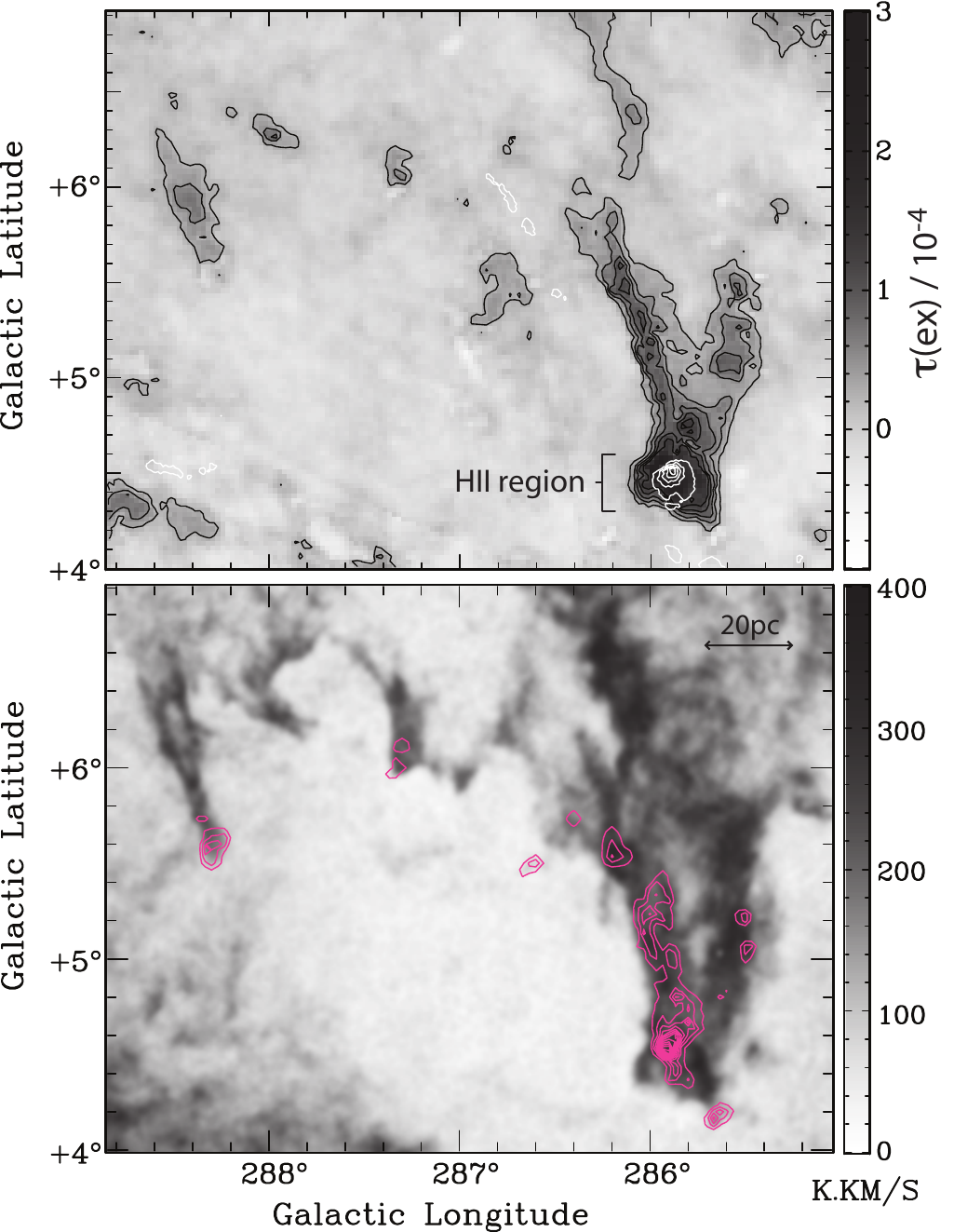}
\caption{Top panel: The greyscale image shows the excess dust opacity at 100 $\mu$m, corresponding to gas not traced in either H{\sc i} or CO. Black contours are drawn at the $5\sigma$ detection level of $2.5\times10^{-5}$, and incremented every $2.5\times10^{-5}$ to a level of $1.5\times10^{-4}$. White contours plot SHASSA H$\alpha$ emission \citep{gaustad01} at a starting level of 550 dR and incremented in steps of 100 dR. Bottom panel: Greyscale H{\sc i} image and CO contours as in panel a1 of figure \ref{fig:287regions}.}
\label{fig:ir_excess}
\end{figure}

\begin{figure}
\epsscale{1.0}
\plotone{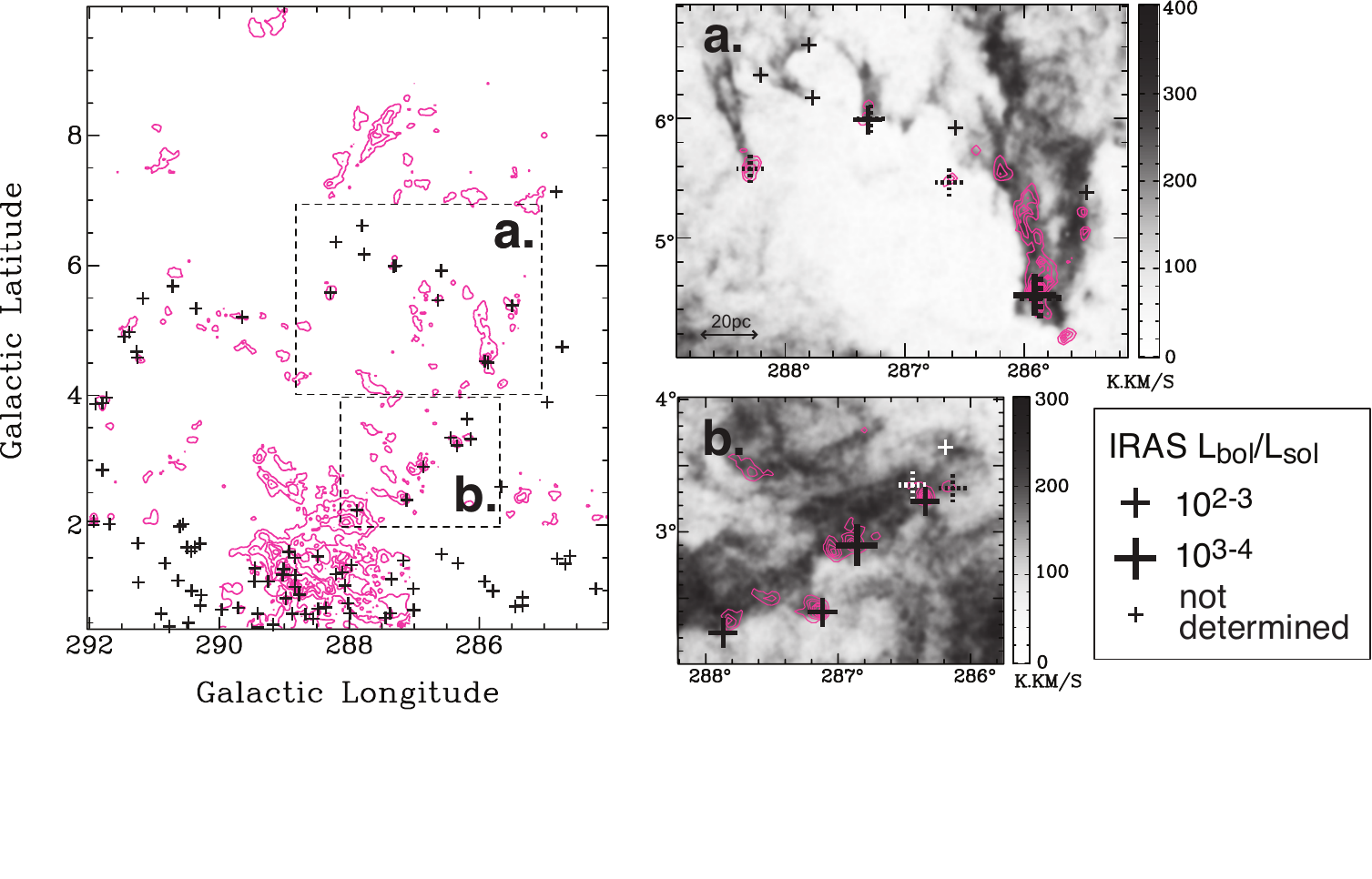}
\caption{IRAS YSO candidates in the GSH 287+04--17 region. The main panel shows $^{12}$CO(J=1--0) integrated over the entire shell velocity range (pink contours) and IRAS YSO candidates (black crosses). Contour levels are $1.5+5.0$ K km s$^{-1}$. Panels (a) and (b) show greyscale H{\sc i} integrated intensity in selected subregions overlaid with CO contours at levels of $1.5+3.0$ K km s$^{-1}$. Velocity integration ranges are (a) $-26.5 < v_{lsr} < -19.9$ km s$^{-1}$ and (b) $-14.1 < v_{lsr} < -10.8$ km s$^{-1}$. IRAS luminosities are calculated only for those objects coincident with molecular gas, for which a genuine association with the shell may therefore be assumed. The IR luminosity in the four IRAS bands is converted to an estimated bolometric luminosity using the correction factor of \citet{myers87}, under the assumption of an SED that peaks at 100 $\mu$m. Filled crosses show those sources for which the 1$\sigma$ uncertainty on $L_{bol}$ is $< 20\%$. Dotted crosses show those for which the uncertainty is greater, or for which the derived luminosity is an upper limit.}
\label{fig:287iraspsc}
\end{figure}

\begin{figure}
\epsscale{1.0}
\plotone{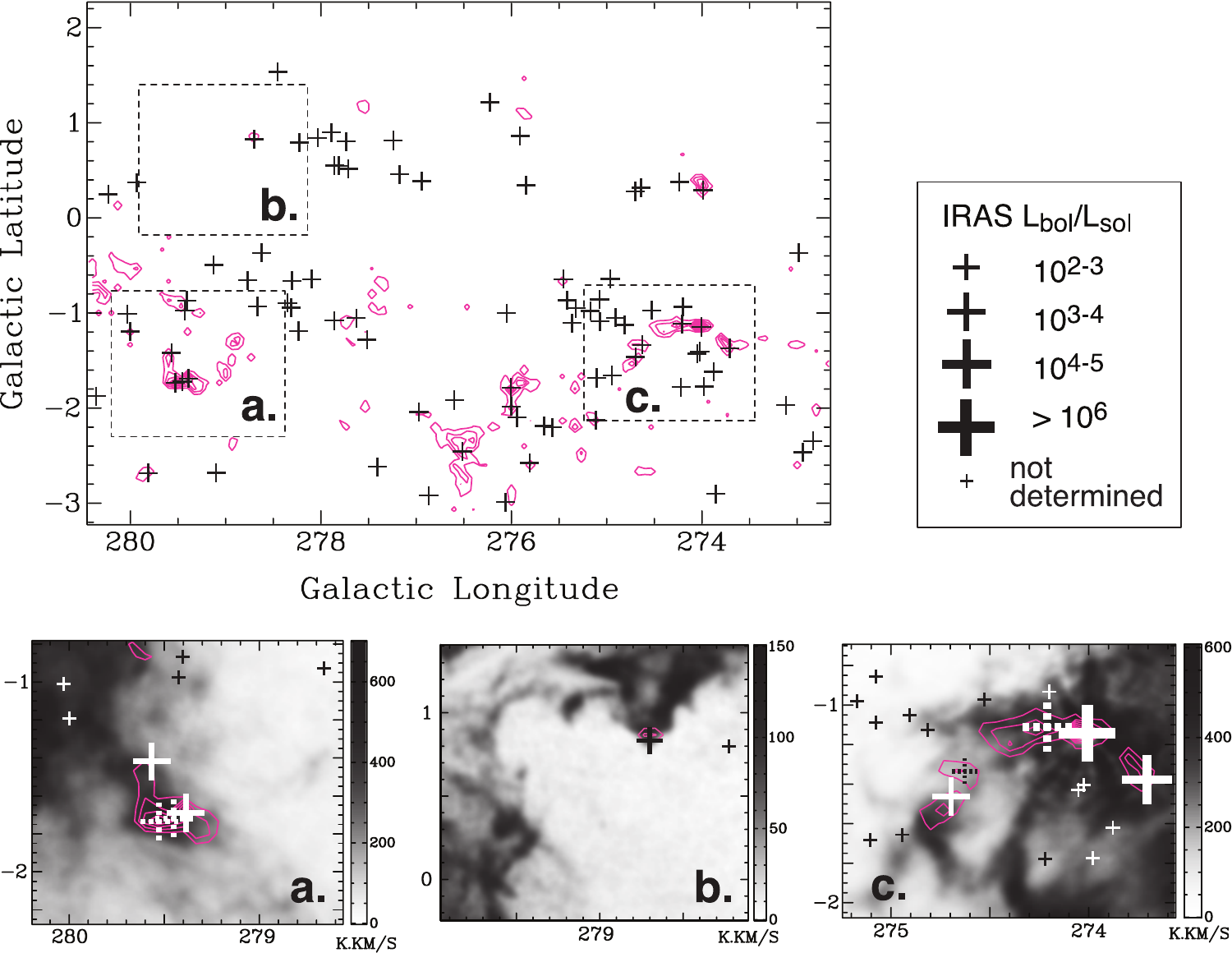}
\caption{IRAS YSO candidates in the GSH 277+00+36 region. The main panel shows $^{12}$CO(J=1--0) integrated over the entire shell velocity range (pink contours) and IRAS YSO candidates (black crosses). Contour levels are $2.0+5.0$ K km s$^{-1}$. Panels (a), (b) and (c) show greyscale H{\sc i} integrated intensity in selected subregions overlaid with CO contours. Velocity integration ranges and contour levels are (a) $40.8 < v_{lsr} < 43.3$ km s$^{-1}$, 4.0+5.0 K km s$^{-1}$, (b) $29.3 < v_{lsr} < 36.7$ km s$^{-1}$, 1.5+1.0 K km s$^{-1}$, and (c) $36.7 < v_{lsr} < 43.3$ km s$^{-1}$, 3.0+5.0 K km s$^{-1}$. IRAS luminosities are calculated only for those objects coincident with molecular gas, for which a genuine association with the shell may therefore be assumed. The IR luminosity in the four IRAS bands is converted to an estimated bolometric luminosity using the correction factor of \citet{myers87}, under the assumption of an SED that peaks at 100 $\mu$m. Filled crosses show those sources for which the 1$\sigma$ uncertainty on $L_{bol}$ is $< 20\%$. Dotted crosses show those for which the uncertainty is greater, or for which the derived luminosity is an upper limit.}
\label{fig:277iraspsc}
\end{figure}

\end{document}